\documentclass[usenatbib]{mn2e_letter}

\usepackage{graphicx}
\usepackage{fixltx2e}
\usepackage{url}

\usepackage{amssymb}
\usepackage{amsmath}

\let\oldhat\hat
\renewcommand{\vec}[1]{\boldsymbol{#1}}
\renewcommand{\hat}[1]{\oldhat{\boldsymbol{#1}}}

\newcommand{\mr}[1]{\mathrm{#1}}

\newcommand{\acknowledgments}{\begin{small}
    \section*{Acknowledgments}\end{small}}
\newcommand\altaffilmark[1]{$^{#1}$}
\newcommand\altaffiltext[1]{$^{#1}$}

\title[Nonlinear Saturation of Buoyancy Instabilities]{Can conduction induce
convection?  The non-linear saturation of buoyancy instabilities in dilute
plasmas}

\author[McCourt, Parrish, Sharma and Quataert]{
\parbox[t]{\textwidth}{
Michael McCourt,\altaffilmark{1}\thanks{E-mail:mkmcc@astro.berkeley.edu}
Ian J. Parrish,\altaffilmark{1,2}
Prateek Sharma\altaffilmark{1,2} \&
Eliot Quataert\altaffilmark{1}}
\vspace*{6pt} \\
\altaffiltext{1}{Department of Astronomy and Theoretical Astrophysics
  Center, University of California
  Berkeley, Berkeley, CA 94720} \\
\altaffiltext{2}{Einstein Fellow} \\
}

\date{Submitted to MNRAS, ?, 2010}
\begin{document}
\maketitle
\label{firstpage}
\begin{abstract}
 We study the effects of anisotropic thermal conduction on low-collisionality,
 astrophysical plasmas using two and three-dimensional magnetohydrodynamic
 simulations.  For weak magnetic fields, dilute plasmas are buoyantly unstable
 for either sign of the temperature gradient: the heat-flux-driven buoyancy
 instability (HBI) operates when the temperature increases with radius while
 the magnetothermal instability (MTI) operates in the opposite limit.  In
 contrast to previous results, we show that, in the presence of a sustained
 temperature gradient, the MTI drives strong turbulence and operates as an
 efficient magnetic dynamo (akin to standard, adiabatic convection).
 Together, the turbulent and magnetic energies contribute up to $\sim$10\% of
 the pressure support in the plasma.  In addition, the MTI drives a large
 convective heat flux, $\sim$~1.5\% $\times \rho c_s^3$.  These findings are
 robust even in the presence of an external source of strong turbulence.  Our
 results on the nonlinear saturation of the HBI are consistent with previous
 studies but we explain physically why the HBI saturates quiescently by
 re-orienting the magnetic field (suppressing the conductive heat flux through
 the plasma), while the MTI saturates by generating sustained turbulence.  We
 also systematically study how an external source of turbulence affects the
 saturation of the HBI: such turbulence can disrupt the HBI only on scales
 where the shearing rate of the turbulence is faster than the growth rate of
 the HBI.  In particular, our results provide a simple mapping between the
 level of turbulence in a plasma and the effective isotropic thermal
 conductivity.  We discuss the astrophysical implications of these findings,
 with a particular focus on the intracluster medium of galaxy clusters.
\end{abstract}


\section{Introduction}
\label{sec:introduction}
 The thermodynamics of a plasma can have dramatic and sometimes unexpected
 implications for its dynamical evolution.  For example, thermal conduction
 can reduce the accretion rate in spherical accretion flows by as much as two
 to three orders of magnitude relative to the Bondi value \citep{Johnson2007,
 Shcherbakov2010}.  More relevant to this paper, \citet{Balbus2000} and
 \citet{Quataert2008} demonstrated that the convective dynamics of conducting
 plasmas are completely different from those of an adiabatic fluid.  This
 paper focuses on the nonlinear evolution and saturation of this convection.

 When anisotropic conduction is rapid compared to the dynamical response of a
 plasma, the temperature gradient, rather than the entropy gradient,
 determines the plasma's convective stability.  The convective instability in
 this limit is known as the heat-flux-driven buoyancy instability (HBI) when
 the temperature increases with height ($\vec{g} \cdot \nabla T < 0$) or the
 magnetothermal instability (MTI) when the temperature decreases with height
 ($\vec{g} \cdot \nabla T > 0$).  We summarize the linear physics of these
 instabilities in section~\ref{sec:background}.  \citet{Parrish2005,
 Parrish2007} and \citet{Parrish2008hbi} studied the nonlinear development of
 the MTI and HBI using numerical simulations.  These instabilities couple the
 magnetic structure of the plasma to its thermal properties and potentially
 have important implications for galaxy clusters \citep{Parrish2008mti,
 Parrish2009, Bogdanovic2009, Sharma2009cr, Parrish2010, Bogdanovic2010,
 Ruszkowski2010}, hot accretion flows onto compact objects \citep{Sharma2008},
 and the interiors and surface layers of white dwarfs and neutron stars
 \citep{Chang2010}.

 In this paper, we revisit the nonlinear behavior of the HBI and MTI.  We
 focus on the physics of their saturation, but also include a lengthy
 discussion of possible astrophysical implications in
 section~\ref{sec:discussion}.  Our analysis is idealized (we use a
 plane-parallel approximation and neglect radiative cooling), but we are able
 to understand the nonlinear behavior of the HBI and MTI, and therefore their
 astrophysical implications, more thoroughly than in previous papers.  Our
 results for the saturation of the HBI are similar to those of
 \citet{Parrish2008hbi}, but with an improved understanding of the saturation
 mechanism.  Our results for the MTI, however, differ from previous results,
 significantly changing the predicted astrophysical implications of the MTI.
 We show in section~\ref{subsec:mti_saturation} that this is because the
 development of the MTI in many previous simulations has been hindered by the
 finite size of the simulation domain.

 The structure of this paper is as follows.  We describe the linear physics of
 the MTI and HBI in section~\ref{sec:background}.  We describe our
 computational setup in section~\ref{sec:method} and the results of our
 numerical simulations in section~\ref{sec:nonlinear_saturation}.  In order to
 better understand how the saturation of the HBI and MTI may change in a more
 realistic astrophysical environment, we study the interaction between these
 instabilities and an external source of turbulence or fluid motion
 (section~\ref{sec:turbulence}).  Finally, in section~\ref{sec:discussion}, we
 summarize our results and discuss the astrophysical implications of our work.

\section{Background}
\label{sec:background}

\subsection{Equations and Assumptions}
\label{subsec:equations}
 We assume that the plasma is an ideal gas with an adiabatic index $\gamma =
 5/3$ and model it using the magneto-hydrodynamic (MHD) equations, neglecting
 all dissipative processes except thermal conduction.  In Cartesian
 coordinates, the equations for the conservation of mass, momentum and
 magnetic flux, and for the evolution of internal energy are
 \begin{subequations}
 \begin{align}
   \frac{\partial \rho}{\partial t} + \nabla \cdot (\rho \, \vec{v})&
      = 0\label{eq:consmass}, \\
   \begin{split}
   \frac{\partial}{\partial t} \left( \rho \, \vec{v} \right)
        + \nabla &\cdot \left[ \rho \, \vec{v}\otimes\vec{v} +
          \left(P + \frac{B^2}{8\pi}\right) \mathbfss{I} +
          \frac{\vec{B} \otimes \vec{B}}{4\pi} \right] \\
      & = \rho \, (\vec{g} + \vec{f}),
   \end{split} \label{eq:consmom} \\
   \vphantom{\frac{\Sigma}{\Sigma}}
   \frac{\partial \vec{B}}{\partial t}&
      = \nabla \times (\vec{v} \times \vec{B})\label{eq:consflux}, \\
   \rho \, T \frac{d s}{d t}& = -\nabla \cdot \vec{Q}_{\mr{cond}}\label{eq:inte},
 \end{align}
 \end{subequations}
 where $\rho$ is the mass density, $\vec{v}$ is the fluid velocity, $\otimes$
 denotes a tensor product, $P$ is the pressure, $\vec{B}$ is the magnetic
 field, $\mathbfss{I}$ is the unit matrix, $\vec{g}$ is the gravitational
 field, $\vec{f}$ is an externally imposed force, $T$ is the temperature,
 \begin{align}
   s = \frac{1}{\gamma - 1}\frac{k_{\mr{B}}}{m_{\mr{H}}}
       \ln\left(\frac{P}{\rho^{\gamma}}\right)
 \end{align}
 is the entropy per unit mass, $d/dt = \partial / \partial t +
 \vec{v}\cdot\nabla$ is the Lagrangian time derivative, and
 $\vec{Q}_{\mr{cond}}$ is the conductive heat flux.

 We ignore the ion contribution to the conductive heat flux, which is smaller
 than the electron contribution by a factor of $(m_{\mr{i}} /
 m_{\mr{e}})^{1/2} \approx 42$.  We assume that the electrons have mean free
 paths much longer than their gyro-radii (as is the case in the intracluster
 medium (ICM) of galaxy clusters); the electrons therefore move almost
 entirely along magnetic field lines.  Consequently, the thermal conductivity
 of the plasma is strongly anisotropic.  The conductive heat flux in this
 limit becomes
 \begin{align}
   \vec{Q}_{\mr{cond}} = -\kappa_{\mr{e}} \, \hat{b} \, 
                        (\hat{b} \cdot \nabla T),
 \end{align}
 where $\hat{b} = \vec{B} / B$ is a unit vector in the direction of the
 magnetic field and $\kappa_{\mr{e}}$ is the thermal conductivity of free
 electrons \citep{Braginskii1965}.  While $\kappa_{\mr{e}}$ depends
 sensitively on temperature \citep{Spitzer1962}, we take it to be constant in
 our calculations to simplify the interpretation of our results.  This
 approximation does not affect our conclusions, because the physics of the
 buoyancy instabilities is independent of the conductivity in the limit that
 the thermal diffusion time across the spatial scales of interest is short
 compared to the dynamical time.

\subsection{The Physics of Buoyancy Instabilities in Dilute Plasmas}
\label{subsec:physics}
 \begin{figure}
    \centering
    \includegraphics[width=0.45\textwidth]{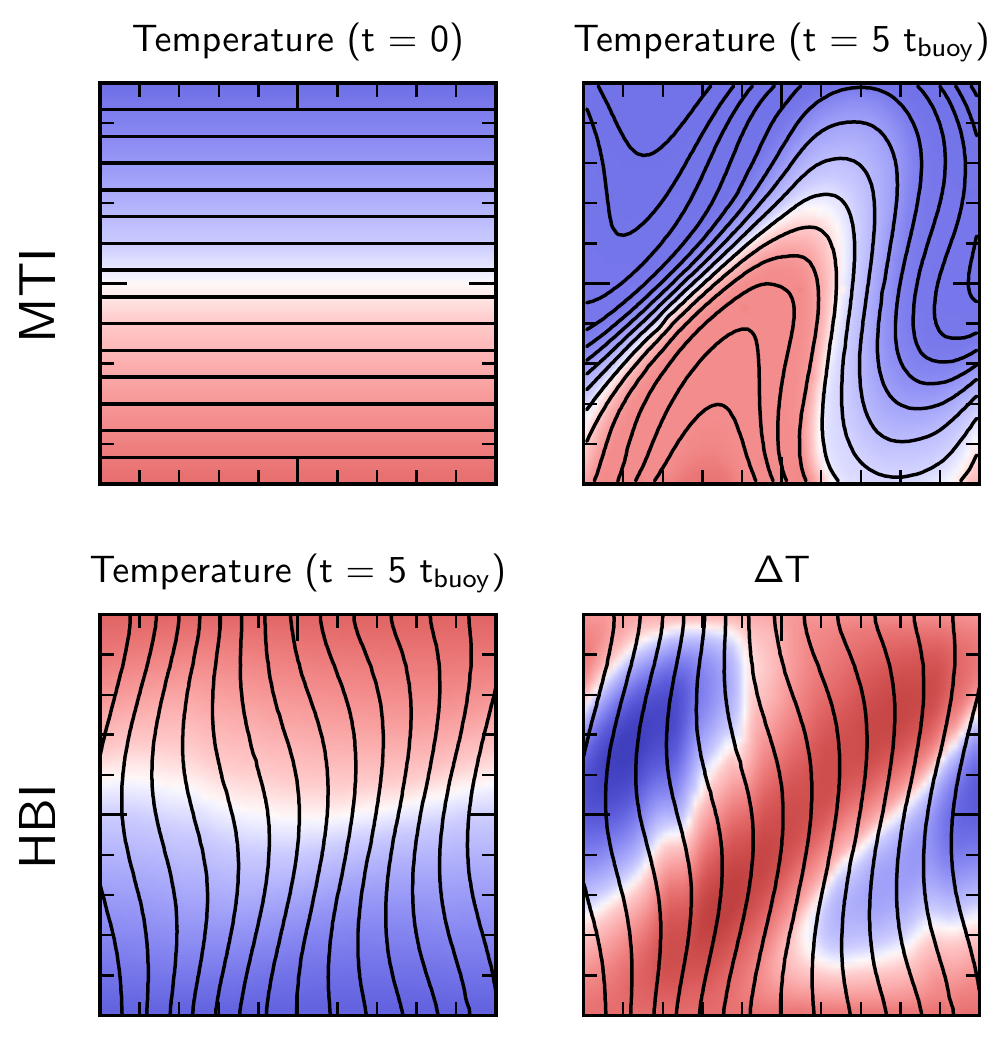}
    \caption{Illustration of the linear development of buoyancy
      instabilities in dilute plasmas from two-dimensional numerical
      simulations.  Color shows the temperature, increasing from blue to red,
      while black lines trace the magnetic field.
      {\bf Top Row:} Quasi-linear evolution of the MTI.  {\em Left panel:}
      Initial equilibrium state, with $\oldhat{b}_z = 0$ and the conductive
      heat flux $\vec{Q} = 0$ everywhere.  {\em Right panel:} The plasma at $t
      = 5 \, t_{\mr{buoy}}$ given an initial perturbation with $\vec{k} = 2
      \pi / L \; \hat{x}$.  Flux freezing and rapid conduction along field
      lines ensure that Lagrangian fluid displacements are nearly isothermal.
      If the initial state has a positive temperature gradient, upwardly
      displaced fluid elements are warmer than their surroundings; they will
      expand and continue to rise.  Similarly, downwardly displaced fluid
      elements are cooler than their surroundings and will sink.
      {\bf Bottom Row:} Quasi-linear evolution of the HBI.  The initial
      equilibrium state has $\oldhat{b}_z = 1$ and a net downward conductive
      heat flux, with $\nabla \cdot \vec{Q} = 0$ everywhere.  {\em Left
      panel:} The plasma at $t = 5 \,t_{\mr{buoy}}$ given an initial
      perturbation with $k_x = k_z$.  The perturbation alters the geometry of
      the magnetic field, so that field lines are not parallel and there can
      be conductive heating and cooling of the plasma. {\em Right panel:}
      Temperature difference at $t = 5 \,t_{\mr{buoy}}$ relative to the
      initial condition. Magnetic field lines converge in upwardly displaced
      fluid elements, leading to an increased temperature.  These fluid
      elements will then expand and continue to rise.  Similarly, downwardly
      displaced fluid elements are conductively cooled, contract and continue
      to sink.}%
      \label{fig:buoyancy_schematic}
 \end{figure}
 The dissipative term in equation~\eqref{eq:inte} shows that fluid
 displacements in a plasma are not in general adiabatic.  The standard
 analysis of buoyancy instabilities therefore does not apply, and the
 convective and mixing properties of a conducting plasma can be very different
 from those of an adiabatic fluid.  In this paper, we focus on the limit in
 which thermal conduction is much faster than the dynamical time in the
 plasma.  This applies on scales $\lesssim 7 \, (\lambda H)^{1/2}$, where
 $\lambda$ is the electron mean free path and $H$ is the plasma scale height;
 this ``rapid conduction limit'' encompasses most scales of interest for the
 ICM.  In this limit, the magnitude of the temperature gradient and the local
 orientation of the magnetic field control the convective stability of the
 plasma \citep{Balbus2000, Quataert2008}.

 Although a single dispersion relation describes the linear stability of
 plasmas in this limit, it is easiest to understand the physics by separately
 considering cases where the temperature increases or decreases with height.
 \citet{Balbus2000} first considered the case where the temperature decreases
 with height and identified the magnetothermal instability, or MTI.  We show a
 schematic of this instability in the top row of
 Figure~\ref{fig:buoyancy_schematic}.  The first panel of this figure shows a
 plasma in hydrostatic and thermal equilibrium, with a weak horizontal
 magnetic field.  We apply a small, plane wave perturbation and, as the plasma
 evolves, the field lines follow the fluid displacements.  Efficient
 conduction along these field lines keeps the displacements isothermal.  Since
 the temperature falls with height, upwardly displaced fluid elements are
 warmer than their new surroundings; they expand and continue to rise.
 Similarly, downwardly displaced fluid elements sink.  An order of magnitude
 calculation shows that displacements grow exponentially, with the growth rate
 (or e-folding rate) $p \sim |g \, \partial \ln T / \partial z|^{-1/2}$.

 \citet{Quataert2008} investigated the limit in which the temperature of the
 plasma increases with height.  The instability in this case is known as the
 heat-flux-driven buoyancy instability, or HBI.  We sketch the growth of the
 HBI in the bottom row of Figure~\ref{fig:buoyancy_schematic}.  Here, the
 initial equilibrium state has vertical magnetic field lines and a constant
 heat flux; the latter is required for the plasma to be in thermal
 equilibrium.  However, perturbations to the magnetic field divert the heat
 flux and conductively heat or cool pockets of the plasma.  If the temperature
 increases with height, upwardly displaced fluid elements become warmer than
 their surroundings and therefore experience a destabilizing buoyant response.
 These perturbations grow with the same growth rate $|g \, \partial \ln T /
 \partial z|^{-1/2}$.

 \citet{Quataert2008} performed a WKB analysis on
 equations~\eqref{eq:consmass}--\eqref{eq:inte} in the Boussinesq limit and
 obtained the following dispersion relation for plasma in a constant
 gravitational field $\vec{g} = -g \, \hat{z}$, threaded by a constant
 magnetic field with any orientation in the $\hat{x}-\hat{z}$~plane:
 \begin{equation}
 \begin{split}
   (p^2 \,+\, &\omega_{\mr{A}}^2) (p + \omega_{\mr{\kappa}}) \;+\; p \, N^2
     (1 - \oldhat{k}_z^2) = \\
   &\omega_{\kappa} \left(g \frac{\partial \ln T}{\partial z} \right)
     \left[ (2 \oldhat{b}_z^2 - 1)(1 - \oldhat{k}_z^2) -
            2 \oldhat{b}_x \oldhat{b}_z \oldhat{k}_x \oldhat{k}_z
     \right]  .\label{eq:dispersion}
 \end{split}
 \end{equation}
 Here, $p = -i \omega$ is the local growth rate of the mode (the time
 dependence of the perturbation is $e^{p t}$), $\omega_{\mr{A}} =
 \vec{k}\cdot\vec{v}_{\mr{A}}$ is the Alfv\'{e}n crossing frequency,
 \begin{align}
   N^2 = \frac{\gamma - 1}{\gamma} \; \frac{\mu \,
   m_{\mr{H}}}{k_{\mr{B}}} \; g \, \frac{\partial s}{\partial z}
 \end{align}
 describes the buoyant response of an adiabatic plasma, $\hat{k} = \vec{k}/k$
 is the direction of the wave vector, and
 \begin{align}
   \omega_{\kappa} = \frac{\gamma - 1}{\gamma} \, \frac{\kappa_{\mr{e}} \, T}{P}
     \, \left(\hat{b} \cdot \vec{k}\right)^2
 \end{align}
 is inversely proportional to the conduction time across the wavelength of the
 mode.  It is convenient to identify $\omega_{\mr{buoy}} = | g \; \partial \ln
 T / \partial z |^{1/2}$ as a characteristic frequency for the buoyancy
 instabilities and we will also use $t_{\mr{buoy}} = \omega_{\mr{buoy}}^{-1}$
 and $t_{\mr{ad}} = N^{-1}$ in our analysis.

 In the limit that conduction is rapid compared to any dynamical response
 ($\omega_{\kappa} \gg \omega_{\mr{A}}, \; N, \; \omega_{\mr{buoy}}$),
 equation~\eqref{eq:dispersion} simplifies to\footnote{In going from the cubic
 equation~\eqref{eq:dispersion} to the
 quadratic~\eqref{eq:simplified_dispersion}, we have ignored a solution for
 $p$.  This solution is exponentially damped on the conduction timescale and
 is not relevant to our analysis.}
 \begin{align}
 \begin{split}
   \left(p^2 + \omega_{\mr{A}}^2 \right) = \mr{sgn}&(\partial T / \partial z)
     \, \omega_{\mr{buoy}}^2 \, \times \\
     &\left[ (2 \oldhat{b}_z^2 - 1)(1 - \oldhat{k}_z^2) -
       2 \oldhat{b}_x \oldhat{b}_z \oldhat{k}_x \oldhat{k}_z
     \right] .\label{eq:simplified_dispersion}
 \end{split}
 \end{align}
 As we stated in section~\ref{subsec:equations}, the growth rate in the rapid
 conduction limit is independent of the thermal conductivity.

 Equation~\eqref{eq:simplified_dispersion} shows that magnetic tension can
 suppress the MTI and HBI; if $\omega_{\mr{A}}^2 > \omega_{\mr{buoy}}^2$,
 $p^2$ must be negative, and the plasma is stable to small perturbations.  In
 this paper, we focus on the relatively weak field limit in which magnetic
 tension does not suppress the instabilities, and we take $\omega_{\mr{A}} \ll
 \omega_{\mr{buoy}}$ in equation~\eqref{eq:simplified_dispersion}.  The
 dominant role of the magnetic field in our analysis is to enforce anisotropic
 electron heat transport.

 For any magnetic field direction $\hat{b}$, the term in square brackets in
 equation~\eqref{eq:simplified_dispersion} can be positive or negative; thus,
 there are always linearly unstable modes, irrespective of the thermal state
 of the plasma (excluding the singular case of $\partial T / \partial z = 0$).

 We finally note that, although we neglected cooling in
 equation~\eqref{eq:inte}, it adds a term to
 equation~\eqref{eq:simplified_dispersion} which is insignificant when the
 cooling time is much longer than $\omega_{\kappa}^{-1}$ \citep{Balbus2008,
 Balbus2010}.  This is not necessarily the case near the centers of cool-core
 clusters, and we plan to explore the combined effects of cooling and buoyancy
 instabilities in future work.  For now, we assume that the buoyancy
 instabilities develop independently of cooling; this allows us to understand
 the nonlinear development and saturation of the instabilities themselves and
 therefore to assess their possible implications for the thermal balance of
 the plasma.

\section{Numerical Method}
\label{sec:method}

\subsection{Problem Setup and Integration}
\label{subsec:setup}
 We consider the evolution of a volume of plasma initially in hydrostatic and
 thermal equilibrium, but subject to either the HBI or MTI.  We seed our
 simulations with Gaussian-random velocity perturbations with a flat spatial
 power spectrum and a standard deviation of $10^{-4} \, c_s$.  The small
 amplitude of these initial perturbations ensures that the instabilities start
 out in a linear phase and permits us to compare our results with the
 predictions of equation~\eqref{eq:simplified_dispersion}.  Astrophysical
 perturbations are unlikely to be this subsonic, however, and the
 instabilities in our simulations take much longer to saturate then one would
 expect from the larger perturbations found in a more realistic scenario.  We
 return to this point in section~\ref{sec:discussion}.

 We highlight the distinctness of the HBI and MTI from adiabatic convection by
 choosing initial conditions with $\partial s / \partial z > 0$ whenever
 possible, so that the plasma would be absolutely stable if it were adiabatic.
 Our results do not rely critically on this choice, however, because the
 plasma is not adiabatic and its evolution is independent of its entropy
 gradient to lowest order in $\omega_{\mr{buoy}} / \omega_{\kappa}$.  Our
 results are much easier to interpret when $\omega_{\mr{buoy}}$ doesn't vary
 across the simulation domain, and we prioritize this constraint on the
 initial condition over the sign of its entropy gradient (as we discuss in
 more detail below).

 We solve equations~\eqref{eq:consmass}--\eqref{eq:consflux}, along with a
 conservative form of equation~\eqref{eq:inte}, using the conservative MHD
 code {\sf Athena} \citep{Gardiner2008, Stone2008} with the anisotropic
 conduction algorithm described in \citet{Parrish2005} and \citet{Sharma2007}.
 In particular, we use the monotonized central difference limiter on
 transverse heat fluxes to ensure stability.  This conduction algorithm is
 sub-cycled with respect to the main integrator with a time step $\Delta t
 \propto (\Delta x)^{2}$; our simulations are therefore more computationally
 expensive than adiabatic MHD calculations.  We draw most of our conclusions
 from simulations performed on uniform Cartesian grids of $(64)^3$ and
 $(128)^3$.  We also performed a large number of two-dimensional simulations
 on grids of $(64)^2$, $(128)^2$ and $(256)^2$.  We found that the kinetic
 energy generated in our local HBI simulations converges by a resolution of
 $(64)^3$; our global HBI simulations require roughly 60 grid cells per scale
 height to give a converged kinetic energy.  The MTI is somewhat more
 sensitive to resolution; we tested the convergence of these simulations by
 comparing the results of an identical simulation performed at $(128)^{3}$ and
 $(256)^{3}$.

 We perform our calculations in the plane-parallel approximation with uniform
 gravity $\vec{g} = - g \, \hat{z}$.  We fix the temperature at the upper and
 lower boundaries of our computational domain, and we extrapolate the pressure
 into the upper and lower ghost cells to ensure that hydrostatic equilibrium
 holds at the boundary.  For all other plasma variables, we apply reflecting
 boundary conditions in the direction parallel to gravity and periodic
 boundary conditions in the orthogonal directions.  As we discuss in
 section~\ref{sec:nonlinear_saturation}, the choice of fixed temperatures at
 the upper and lower boundaries has an important effect on the non-linear
 evolution.  Simulations with Neumann boundary conditions in which the
 temperature at the boundaries is free to adjust would give somewhat different
 results (see, e.g., \citealt{Parrish2005}).  Our choice of Dirichlet boundary
 conditions is largely motivated by the fact that many galaxy clusters in the
 local universe are observed to have non-negligible temperature gradients
 \citep{Piffaretti2005}.

 We perform both local (with the size of the simulation domain $L$ much
 smaller than the scale height $H$) and global ($L \gtrsim H$) simulations of
 the MTI and HBI.  The local simulations separate the development of the
 instability from any large-scale response of the plasma, allowing us to study
 the dynamics in great detail.  However, because the response of the plasma on
 larger scales can influence the nonlinear evolution and saturation of the
 instabilities, we also carry out global simulations.

\subsection{Local Simulations}
\label{subsec:local_setup}
 In sections~\ref{sec:method}--\ref{sec:turbulence}, we work in units with
 $k_{\mr{B}} = \mu \, m_{\mr{p}} = 1$.  As noted previously, we restrict our
 analysis to plasmas with rapid conduction; we find that setting
 $\kappa_{\mr{e}} = 10 \times \rho \, \omega_{\mr{buoy}} \, L^2$ puts our
 local simulations safely in this limit.  This corresponds to a thermal
 diffusion time across the box of $\sim 0.1 \, \omega_{\mr{buoy}}^{-1}$.  We
 initialize our local HBI simulations with a linear temperature gradient:
 \begin{subequations}
 \begin{align}
   T(z)& = T_0 \left( 1 + z / H \right), \\
   \rho(z)& = \rho_0 \left( 1 + z / H \right)^{-3}, \\
   P(z)& = \rho(z) \; T(z) \; .
 \end{align}\label{eq:hbi_local_setup}
 \end{subequations}
 We choose to set $\rho_0 = 1$ and $g = 1 = 2 \; T_0 / H$.  Unless otherwise
 noted, we take $H = 2$ ($T_0 = 1$) in our local simulations and we evolve a
 volume of plasma from $z = 0$ to $z = L = 0.1$, i.e., over $\sim 5\%$ of a
 scale-height.

 We use the setup of \citet{Parrish2005} for our local MTI simulations:
 \begin{subequations}
 \begin{align}
   T(z)& = T_0 \left( 1 - z / H \right), \\
   \rho(z)& = \rho_0 \left( 1 - z / H \right)^{2}, \\
   P(z)& = \rho(z) \; T(z),
 \end{align}\label{eq:mti_local_setup}
 \end{subequations}
 with $g = 1$ and $H = 3$.  The MTI induces large vertical displacements in
 the plasma, and our reflecting boundary conditions clearly influence its
 evolution; we attempt to minimize the effects of the boundaries by
 sandwiching the unstable volume of plasma between two buoyantly neutral
 layers.  \citet{Parrish2005} describe this setup in more
 detail.\footnote{Note that, although \citet{Parrish2005} describe these extra
 layers as buoyantly stable, the effect of the isotropic conductivity is to
 make them buoyantly neutral.}

 Both of the above atmospheres have positive entropy gradients and therefore
 would be stable in the absence of anisotropic conduction.  The results we
 describe in this paper are entirely due to non-adiabatic processes.

 We show in section~\ref{subsec:hbi_saturation} that local and global HBI
 simulations give very similar results; we therefore use the simpler, local
 simulations for most of our analysis of the HBI.  By contrast, the
 large-scale vertical motions induced by the MTI make its evolution inherently
 global; as we describe in section~\ref{subsec:mti_saturation}, local
 simulations do not give a converged result independent of $L / H$.  We
 therefore study the saturated state of the MTI using global simulations,
 which require an alternate set of initial conditions.

\subsection{Global Simulations}
\label{subsec:global_setup}
 The physical properties of the plasma in a global simulation can vary by an
 order of magnitude or more across the simulation domain.  This complicates
 our study of the HBI and MTI because the instabilities can be in different
 stages of evolution at different spatial locations.  Similarly, the kinetic
 and magnetic energies generated by the instabilities in a global simulation
 can be functions of height, obscuring their dependence on other parameters.
 While one must confront these problems when studying buoyancy instabilities
 in an astrophysical context, we avoid them by choosing initial conditions
 with a buoyancy time that is constant with height.  We specify
 \begin{subequations}
 \begin{align}
   g(z)& = g_0 \; e^{-z/S}, \\
   T(z)& = \exp \left[ \pm \frac{S \omega^2_{\mr{buoy}}}{g_0}
                       \left(e^{z/S} - 1 \right)\right]  .  \label{eq:global_T}
 \end{align}\label{eq:global_setup}
 \end{subequations}
 The positive and negative signs in equation~\eqref{eq:global_T} produce
 atmospheres unstable to the HBI and MTI, respectively.  Note that $T(z=0) =
 1$ as in our local simulations.  We take $g_0 = 1$, $S = 3$ and
 $\omega^2_{\mr{buoy}} = 1/2$ and numerically solve for $\rho(z)$ so that the
 initial atmosphere is in hydrostatic equilibrium.  The atmosphere defined by
 equation~\eqref{eq:global_T} does not have a single, well-defined scale
 height.  We therefore define $H$ so that $L/H = \ln\left(T_{\mr{max}} /
 T_{\mr{min}}\right)$; this definition for $H$ is analogous to that used in
 our local simulations and $L/H$ reflects the total free energy available to
 the instabilities.  Because $H$ is a defined, rather than fundamental,
 property of the atmosphere, it is not independent of the size of the
 simulation domain.  We carry out simulations with domain sizes $L$ = 0.5 and
 2.0, which have $L/H$ = 0.27 and 1.4, respectively.

 Simulations with a larger simulation domain size inherently permit larger
 vertical displacements, and the boundary conditions do not influence the
 evolution of the plasma as strongly as they do in a local simulation.  We
 find that the neutrally stable layers described in
 section~\ref{subsec:local_setup} don't alter the results of our global
 simulations, and thus we do not include them in our setup.

 We chose the conductivity in our local simulations so that $\kappa_{\mr{e}} =
 10 \, \rho \, \omega_{\mr{buoy}} L^2$, but the corresponding constraint on
 the time step becomes impractical for our global simulations.  Instead, we
 adjust $\kappa_{\mr{e}}$ so that the ratio $\kappa_{\mr{e}} / L$ is the same
 as it is in the local simulations.  This keeps the ratio of the conduction
 time to the sound crossing time constant, and is appropriate if the
 turbulence driven by the MTI or HBI reaches a terminal speed less than or of
 order the sound speed.  We carried out simulations with different values of
 the conductivity and verified that our results are insensitive to factor of
 few changes in $\kappa_{\mr{e}}$.

 The atmosphere we use for our global MTI simulations has a negative entropy
 gradient. One might worry that the results of these simulations---which aim
 to focus on the MTI---would be biased by the presence of adiabatic
 convection.  This is, however, only a pedagogical inconvenience;
 $\omega_{\kappa} > N$ on all relevant scales in the simulation, so the
 adiabatic limit to equation~\eqref{eq:dispersion} is not important, and the
 evolution of the plasma is nearly independent of its entropy gradient.  To
 confirm this we carried out $L = 0.27 \, H$ simulations with both sets of
 initial conditions; they give very similar results.  Additionally, we
 performed one simulation of adiabatic convection using this setup, and the
 behavior of this simulation is entirely different from the MTI, especially at
 late times.

\subsection{Turbulence}
\label{subsec:turbulence}
 The pure HBI/MTI simulations described in the previous sections are
 physically very instructive but astrophysically somewhat idealized.  In order
 to better understand the astrophysical role of the HBI and MTI, we also study
 their interaction with other sources of turbulence and fluid motion.  We
 assume that these take the form of isotropic turbulence; we include such
 turbulence via the externally imposed force field $\vec{f}$ in
 equation~\eqref{eq:consmom}.  Following \citet{Lemaster2008}, we compute
 $\vec{f}$ in momentum space from scales $k_0 = \left\{4, 6, 8\right\} \times
 2 \pi / L$, down to $k_{\mr{max}} = 2 \, k_0$, with an injected energy
 spectrum $\dot{E}_{k}^{\mathrm{(inj)}} \propto k^{-3}$. We then randomize the
 phases, perform a Helmholtz decomposition of the field, discard the
 compressive component, transform the field into configuration space, and
 normalize it to a specified energy injection rate. We find that the
 turbulence sets up a nonlinear cascade that is not very sensitive to either
 the driving scale or the injected spectrum of the turbulence.

 This prescription for turbulence is statistically uniform in both space and
 time and therefore provides a controlled environment in which to study the
 interaction between turbulence and buoyancy instabilities.  This may not,
 however, be a good approximation to real astrophysical turbulence; we discuss
 the implications of our choice in sections~\ref{sec:turbulence}
 and~\ref{sec:discussion}.

\section{Nonlinear Saturation}
\label{sec:nonlinear_saturation}
 \begin{figure*}
   \centering
   \includegraphics[width=0.9\textwidth]{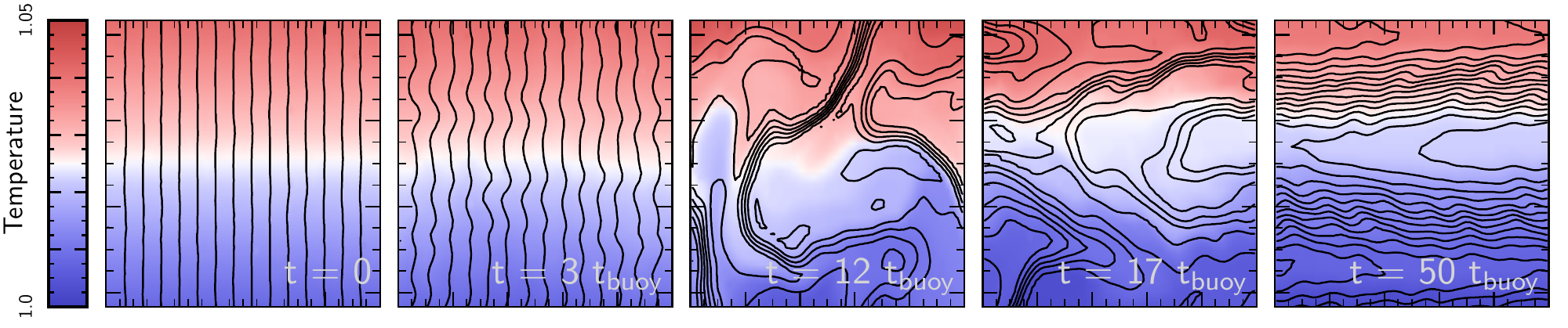}
   \caption{Evolution of the HBI with an initially vertical magnetic field in
     a local, 2D simulation (simulation~h1 in Table~\ref{tab:hbi_pure}).
     Color shows temperature and black lines show magnetic field lines.  A
     small velocity perturbation to the initial state seeds exponentially
     growing modes which dramatically reorient the magnetic field to be
     predominantly horizontal.  The induced velocities are always highly
     subsonic and, after $t \sim 20 \, t_{\mr{buoy}}$, are also almost
     entirely horizontal.  Once the plasma reaches its saturated state, it is
     buoyantly stable to vertical displacements.  The plasma does not resist
     horizontal displacements, but the saturated state is nearly symmetric to
     these displacements and they do not change its character.}
   \label{fig:hbi_5panel}
  \end{figure*}  

\subsection{Saturation of the HBI}
\label{subsec:hbi_saturation}
 \citet{Parrish2008hbi} described the nonlinear saturation of the HBI, but
 they did not explicitly test the dependence of their results on the size of
 the computational domain.  This dependence turns out to be crucial for the
 MTI (see \S~\ref{subsec:mti_saturation}), but we show here that the size of
 the domain has little effect on the saturation of the HBI.  Nonetheless, we
 describe the nonlinear behavior of the HBI in reasonable detail, expanding on
 the physical interpretation given in previous papers (although our HBI
 results do not differ significantly from those of previous authors).  The
 saturation of the HBI is the simplest process we consider in this paper and
 serves as a useful comparison for our new results.

 \begin{table}
   \caption{Parameters for the HBI simulations
     (\S~\ref{subsec:hbi_saturation}).}
   \label{tab:hbi_pure}
   \begin{center}
     \begin{tabular}{@{}ccccc}
       \hline
       Name & $D$ & res & $L/H$ & $\kappa$ \\
       \hline
         h1 &   2 &  64 &  0.05 &     7.07 \\ 
         h2 &   3 &  64 &  0.05 &     7.07 \\ 
         h3 &   3 & 128 &  0.75 &     0.47 \\ 
         h4 &   3 & 128 &  1.40 &     0.35 \\ 
       \hline
     \end{tabular}
   \end{center}

   \medskip
   All simulations are performed on square Cartesian grids of size $L$.  $D$
   is the dimensionality of the simulation, res is the number of grid cells
   along a side, and $\kappa$ is the conductivity (in units of $k_{\mathrm{B}}
   / \mu m \times \rho \, \omega_{\mr{buoy}} L^2 $).  All of these simulations
   are local (eq.~\ref{eq:hbi_local_setup}), except for the one with $L/H =
   1.4$, which uses the global setup (eq.~\ref{eq:global_setup}).  We
   initialized all of these simulations with weak horizontal magnetic fields
   ($B/\sqrt{4\pi} = 10^{-6}$).
 \end{table}

 We study the saturation of the HBI using 2D and 3D simulations spanning a
 range of domain sizes $L/H$.  Table~\ref{tab:hbi_pure} lists all of the
 simulations presented in this section.

\begin{figure}
   \centering
   \includegraphics[width=0.45\textwidth]{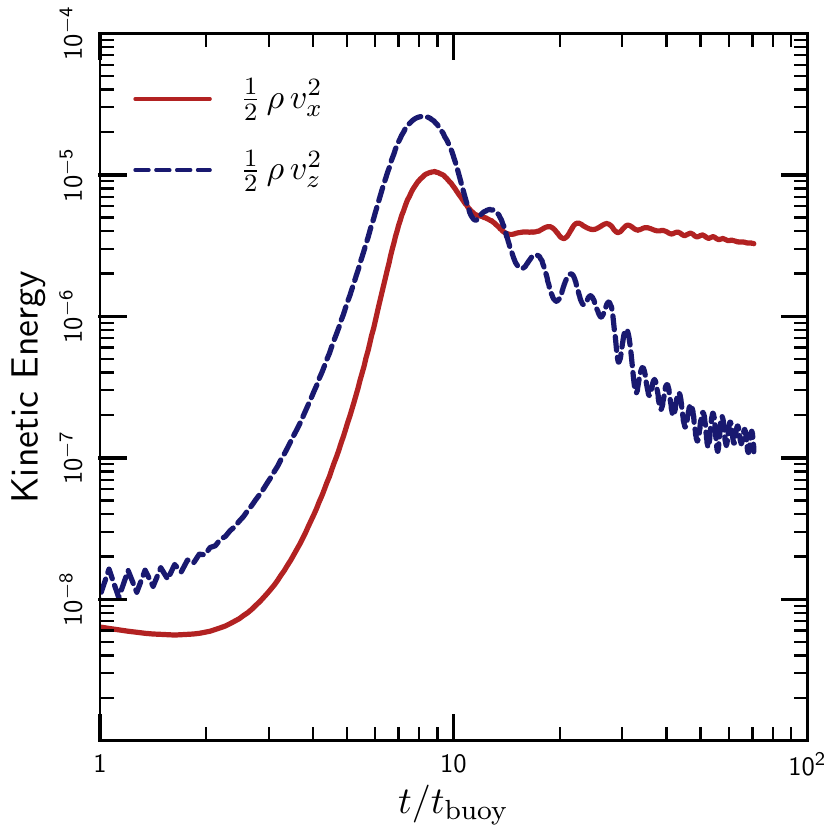}
   \caption{Evolution of the vertical and horizontal kinetic energy in
     a local, 2D HBI simulation (simulation h1 in Table~\ref{tab:hbi_pure}).
     The units are such that the thermal pressure $P \approx 1$ and the
     initial magnetic energy is $B^2 / 8\pi = 10^{-12}$.  After a period of
     exponential growth in which the $x$ and $z$ motions are in approximate
     equipartition, the HBI saturates and the kinetic energy ceases to grow.
     At this point, the energy in the vertical motion is in the form of stable
     oscillations, which decay non-linearly.  The horizontal motions are
     unhindered, however, and persist for the entire duration of the
     simulation.  These horizontal motions are responsible for the asymmetry
     of the magnetic field shown in Figure~\ref{fig:hbi_magnetic}.}%
   \label{fig:hbi_ke_split}
 \end{figure}

 Figure~\ref{fig:hbi_5panel} shows snapshots of the evolution of temperature
 and magnetic field lines in a local, 2D HBI simulation.  We chose this
 simulation to simplify the field-line visualization, but the results in
 Figure~\ref{fig:hbi_5panel} apply equally to our local and global 3D
 simulations.  We initialized this simulation in an unstable equilibrium state
 with vertical magnetic field lines ($\oldhat{b}_z = 1$).  As described in
 section~\ref{subsec:setup}, we seed this initial condition with small
 velocity perturbations; the HBI causes these perturbations to grow in the
 first three panels of Figure~\ref{fig:hbi_5panel}.  The evolution becomes
 nonlinear in the third panel, when the velocity perturbations reach $\sim
 4\%$ of the sound speed.  Afterwards, the instability begins to saturate and
 the plasma slowly settles into a new equilibrium state.  The last panel in
 Figure~\ref{fig:hbi_5panel} shows that this saturated state is highly
 anisotropic: the magnetic field lines are almost entirely orthogonal to
 gravity.  Flux conservation implies that the fluid motions must also be
 anisotropic, with most of the kinetic energy in horizontal motions at late
 times (see Fig.~\ref{fig:hbi_ke_split}, discussed below).  These horizontal
 motions are very subsonic: in all of our simulations, the velocities
 generated by the HBI are significantly less than $1\%$ of the sound speed in
 the saturated state.

 Because the fluid velocities remain small, the linear dispersion relation
 (eq. ~\ref{eq:simplified_dispersion}) captures much of the evolution of the
 HBI, even at late times.  For any magnetic field orientation, the fastest
 growing modes are the ones with $\vec{k}$ along the axis $\hat{b} \times
 (\hat{b} \times \vec{g})$; these modes have the growth rate
 \begin{align}
   p_{\mr{max}} = |\omega_{\mr{buoy}} \, \oldhat{b}_z | ,
   \label{eq:hbi_max_growth}
 \end{align}
 which decreases as the field lines become horizontal.  Additionally, when
 $\oldhat{b}_{z}^{2} < 1/2$, only modes with $\oldhat{k}_{z}^{2}
 > 1-4(\oldhat{b}_z^2 - \oldhat{b}_z^4)$ are unstable.  Since the HBI
 saturates by making the field lines horizontal ($\oldhat{b}_z \rightarrow
 0$), both the maximum growth rate of the instability and the volume of phase
 space for unstable modes decrease as the HBI develops.  This strongly limits
 the growth of the perturbations, and helps explain why the instability
 saturates relatively quiescently.

 As argued by \citet{Parrish2008hbi}, the HBI saturates when its maximum
 growth rate $p_{\mr{max}}$ vanishes, so that there are no longer unstable
 modes.  While this is clearly a sufficient condition for the plasma to reach
 a new stable equilibrium, it is by no means necessary: the instability could,
 e.g., saturate via nonlinear effects, but in practice this is not the case
 (at least for simulations without an additional source of turbulence).
 Equation~\eqref{eq:hbi_max_growth} for $p_{\mr{max}}$ shows that the HBI
 could saturate by making either $\partial T / \partial z$ or $\oldhat{b}_z$
 vanish; intuitively, the HBI is powered by a conductive heat flux, which it
 must extinguish in order to stop growing.  Erasing the temperature gradient
 might seem like the more natural saturation channel, since the conduction
 time across the domain is much shorter than the time it takes the HBI to
 develop and saturate.  In an astrophysical setting, however, the large-scale
 temperature field is often controlled by cooling, accretion or other
 processes apart from the HBI.  We therefore impose the overall temperature
 gradient in our simulations by fixing the temperature at the top and bottom
 of the domain, so that $\omega_{\mr{buoy}}$ is roughly independent of time
 and saturation requires $\oldhat{b}_z = 0$.

 Since the HBI saturates by making the magnetic field lines horizontal, we
 take the $\oldhat{b}_z \rightarrow 0$ limit in
 equation~\eqref{eq:simplified_dispersion} to understand the late-time
 behavior of the plasma:
 \begin{align}
   \omega = \pm \, \omega_{\mr{buoy}} \left(1-\oldhat{k}_z^2\right)^{1/2} .
   \label{eq:hbi_stable_osc}
 \end{align}
 We have assumed here that the magnetic field is weak enough for magnetic
 tension to be negligible on the scales of interest.\footnote{Note that
 equation~\ref{eq:hbi_stable_osc} only strictly applies when the magnetic
 field is {\it exactly} horizontal.  More generally, there will still be
 unstable modes with growth rate given by equation~\ref{eq:hbi_max_growth};
 this growth rate is very slow in the saturated state of the HBI, however, and
 these modes don't change the dynamics of the plasma.}
 Equation~\eqref{eq:hbi_stable_osc} shows that the saturated state of the HBI
 is buoyantly stable, but that there is a family modes with $\oldhat{k}_z = 1$
 which feel no restoring force.  This simply reflects the fact that the plasma
 is stably stratified and resists vertical displacements.  Displacements
 orthogonal to gravity are unaffected by buoyancy, however, and appear as
 zero-frequency modes in the dispersion relation.

 Figure~\ref{fig:hbi_ke_split} demonstrates this asymmetry between vertical
 and horizontal displacements in the late time evolution of the plasma.  This
 figure shows the kinetic energy in horizontal and vertical motions as a
 function of time.  During the initial, linear growth of the instability ($t
 \lesssim 10 \, t_{\mr{buoy}}$), buoyantly unstable fluid elements accelerate
 toward the stable equilibrium, and the kinetic energy is approximately evenly
 split among vertical and horizontal motions.  As the instability saturates,
 however, the plasma becomes buoyantly stable and traps the vertical motions
 in decaying oscillations (internal gravity waves).  The horizontal motions
 keep going, however, and retain their kinetic energy for the duration of the
 simulation.  This difference in the response of the plasma to vertical and
 horizontal motions accounts for the anisotropy of the velocity field in the
 saturated state of the HBI.

\begin{figure}
   \centering
   \includegraphics[width=0.45\textwidth]{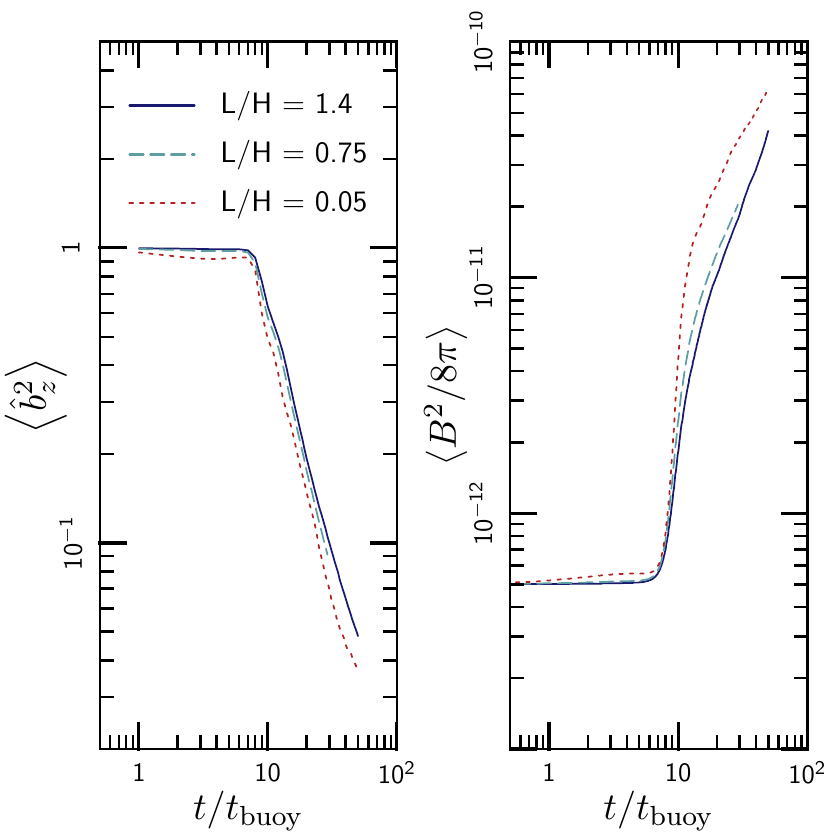}
   \caption{Evolution of the orientation (left) and energy (right) of
     the magnetic field in 3D HBI simulations for three different values of
     the size of simulation domain relative to the temperature scale-height
     ($L/H$) (simulations h2--h4 in Table~\ref{tab:hbi_pure}).  Units are such
     that the thermal pressure $P$ is $\simeq 1$.  The results are nearly
     independent of size of the simulation domain.  As described in
     \S~\ref{subsec:hbi_saturation}, the HBI saturates by shutting itself off;
     the linear exponential growth ends at $t \sim 10 \, t_{\mr{buoy}}$, and
     most of the evolution of the magnetic field happens afterward.  This
     evolution is driven by the horizontal motions shown in
     Figure~\ref{fig:hbi_ke_split}, which both amplify and reorient the
     magnetic field.  After a brief period of exponential growth, the field
     amplification is roughly linear in time.  By contrast, the field
     amplification by the MTI is exponential in time (see
     Fig.~\ref{fig:mti_magnetic}). }%
   \label{fig:hbi_magnetic}
 \end{figure}

 Figure~\ref{fig:hbi_magnetic} shows the evolution of the rms magnetic field
 angle (left panel) and magnetic energy (right panel) in 3D HBI simulations
 for three different values of the size of the computational domain $L$
 relative to the scale height $H$. The zero frequency modes discussed above
 also dominate the evolution of the magnetic field at late times, after the
 motions become nonlinear ($t \gtrsim 10 \, t_{\mr{buoy}}$).  The horizontal
 displacements stretch out the field lines, amplifying and reorienting them.
 Quantitatively, we expect that $\oldhat{b}_z \sim \lambda / \xi \propto
 t^{-1}$, where $\lambda$ is a characteristic scale for the modes in the
 saturated state, $\xi$ is the magnitude of the horizontal displacements, and
 we have assumed that the velocity is constant with time.  The left panel of
 Figure~\ref{fig:hbi_magnetic} shows that the dependence in the simulations is
 quite close to this, with $\oldhat{b}_z \propto t^{-0.85}$.\footnote{The
 slight difference relative to the simple predictions of flux freezing given
 the velocity field in Figure~\ref{fig:hbi_ke_split} may be due to the finite
 resolution of our simulations, which prevents us from resolving the field
 line direction when $\oldhat{b}_z \lesssim 10 \times dz / L$.}  Stretching
 the field lines in this manner amplifies the field strength by an amount
 $\delta B \propto \xi$; if the velocity is constant with time, we expect $B^2
 \propto t^2$.  The right panel of Figure~\ref{fig:hbi_magnetic} shows that
 this time dependence is approximately true for our two larger HBI
 simulations; the amplification is slightly slower in the very local
 calculation with $L/H = 0.05$.  There is no indication that the magnetic
 field amplification has saturated at late times in the HBI simulations.  We
 suspect that the amplification would continue until the magnetic and kinetic
 energy densities reach approximate equipartition, but we would have to run
 the simulation for a very long time to verify this.

 One of the important results in Figure~\ref{fig:hbi_magnetic} is that the
 saturated state of the HBI is nearly independent of the size of the
 computational domain $L/H$.  Since the late time evolution of the plasma is
 driven only by horizontal displacements, the key dynamics all occur at
 approximately the same height in the atmosphere. The saturation of the HBI is
 thus essentially local in nature and should not be sensitive to the global
 thermal state of the plasma or the details of the computational setup.

 %

 The dramatic reorienting of the magnetic field caused by the HBI severely
 suppresses the conductive heat flux through the plasma.  The conductive flux
 is proportional to $\langle \oldhat{b}_z^2 \rangle$, which decreases in time
 $\propto (t/t_{\mr{buoy}})^{-1.7}$.  The saturated state of the HBI is also
 buoyantly stable and resists any vertical mixing of the plasma.  As a result,
 the convective energy fluxes in our HBI simulations are very small, $\sim
 10^{-6} \, \rho c_s^3$.  The effect of the HBI therefore is to strongly
 insulate the plasma against both conductive and convective energy transport.
 This can dramatically affect the thermal evolution of the plasma
 \citep{Parrish2009, Bogdanovic2009}.

 The fact that the growth rate of the HBI depends on the local orientation of
 the magnetic field, as well as the thermal structure of the plasma, makes it
 very different from adiabatic convection.  This dependence on the magnetic
 field structure provides a saturation channel in which the kinetic energies
 are very small compared to the thermal energy (e.g., in
 Fig.~\ref{fig:hbi_ke_split}, $\rho v^2 / n k T \sim 10^{-5}$).  Critically,
 these highly subsonic motions occur in simulations in which the boundary
 conditions allow for the presence of a sustained, order unity, temperature
 gradient ($d\ln T/d\ln z \sim 1$).  In an adiabatic simulation, the analogous
 sustained entropy gradient would generate convective motions with $\rho v^2
 \sim n k T$.  This does not occur in an HBI-unstable plasma.  Thus, although
 a plasma with a positive temperature gradient is in general buoyantly
 unstable, the effect of the HBI is to peacefully stabilize the plasma within
 a few buoyancy times by suppressing the conductive heat flux through the
 plasma.  The resulting, stably-stratified plasma then resists vertical mixing
 and, in the absence of strong external forcing, we expect the fluid
 velocities and magnetic field lines to be primarily horizontal.  In
 section~\ref{sec:turbulence}, we perturb this state with externally driven,
 isotropic turbulence and test the strength of the stabilizing force.

\subsection{Saturation of the MTI}
\label{subsec:mti_saturation}
 \begin{figure*}
   \includegraphics[width=0.9\textwidth]{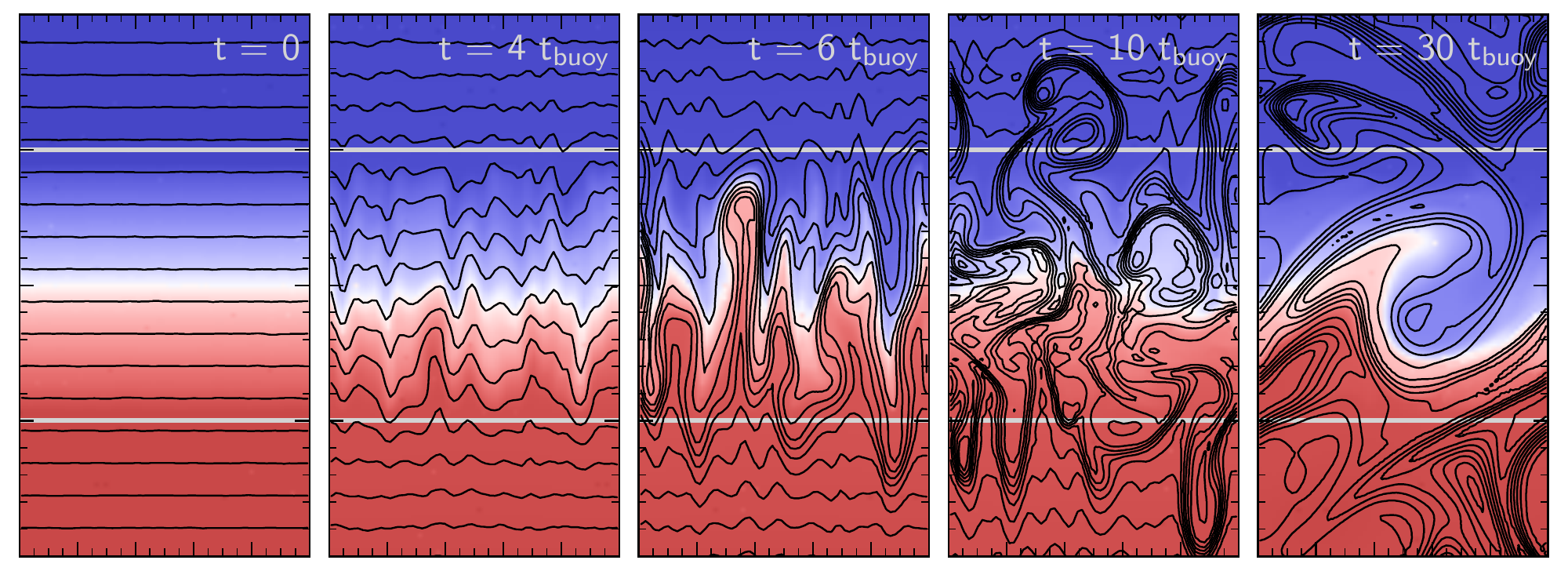}
   \caption{Evolution of the MTI with an initially horizontal magnetic
     field in a local, 2D simulation (simulation~m1 in
     Table~\ref{tab:mti_pure}).  Gray horizontal lines show the transition to
     the buoyantly neutral layers described in \S~\ref{subsec:local_setup};
     the color scale is identical to that in Figure~\ref{fig:hbi_5panel}.
     Initial perturbations grow by the mechanism described in
     \S~\ref{subsec:physics} (Fig. \ref{fig:buoyancy_schematic}); rising and
     sinking plumes rake out the field lines until, by $t = 6 \,
     t_{\mr{buoy}}$, they are mostly vertical.  This configuration is,
     however, nonlinearly unstable to horizontal displacements, which generate
     a horizontal magnetic field and thus continually seed the MTI (see
     Fig.~\ref{fig:mti_5panel_v}).  The result is vigorous, sustained
     convection in marked contrast to the saturation of the HBI in
     Fig.~\ref{fig:hbi_5panel}.  In this local simulation, buoyant plumes
     accelerate until they reach the neutrally stable layers.  The boundaries
     prematurely stop the growth of the MTI, and the local simulation under
     predicts the kinetic energy generated by the MTI (see
     Fig. \ref{fig:mti_kinetic}).}%
   \label{fig:mti_5panel_h}
   \centering
   \includegraphics[width=0.9\textwidth]{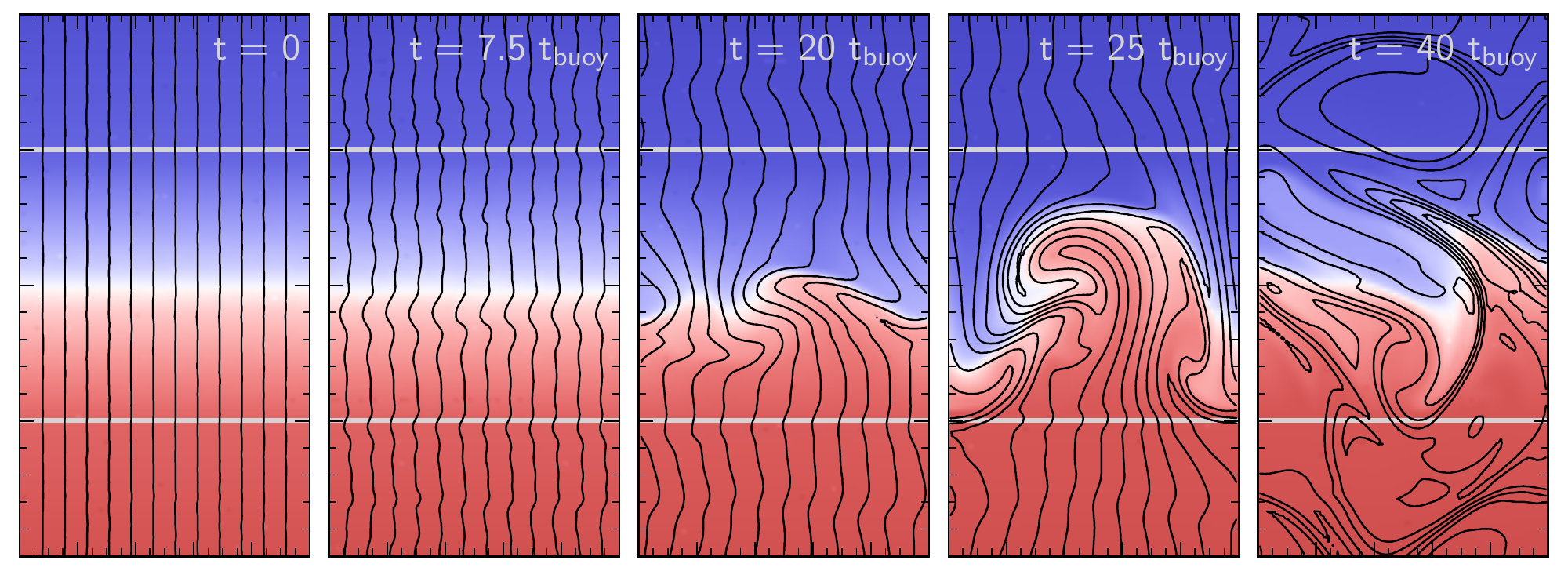}
   \caption{Evolution of the MTI in a plasma with an initially
     vertical magnetic field in a local, 2D simulation (simulation~m2 in
     Table~\ref{tab:mti_pure}).  This configuration is linearly stable
     according to equation~\eqref{eq:simplified_dispersion}, but there are
     zero-frequency, $k_x = 0$, modes which do not have a restoring force.
     Physically, these correspond to horizontal motions which do not feel
     gravity/buoyancy.  Because of this zero frequency mode, small random
     initial perturbations add a horizontal component to the magnetic field,
     eventually rendering the plasma unstable to the MTI. This creates a
     feedback loop, allowing the MTI to generate vigorous, sustained
     convection; the HBI does not have this same feedback loop and so does not
     generate sustained turbulence (Fig.~\ref{fig:hbi_5panel}).  At late
     times, the results of this simulation with an initially vertical magnetic
     field are very similar to Figure~\ref{fig:mti_5panel_h} which starts with
     a horizontal magnetic field.}%
   \label{fig:mti_5panel_v}
 \end{figure*}
 Figure~\ref{fig:mti_5panel_h} shows the evolution of one of our local, 2D MTI
 simulations.  As in the HBI simulation shown in Figure~\ref{fig:hbi_5panel},
 we initialized this simulation in an unstable equilibrium state (a weak
 horizontal magnetic field) and seeded it with the small velocity
 perturbations described in section~\ref{subsec:setup}.  The MTI and HBI stem
 from very similar physics, and as a result have very similar linear dynamics.
 The nonlinear behavior of the two instabilities is entirely different,
 however.  While the HBI saturates relatively quiescently by driving the
 plasma to a buoyantly stable and highly anisotropic state, the MTI generates
 vigorous, sustained convection that tends to isotropize both the magnetic and
 velocity fields.

 \begin{table}
   \caption{Parameters for the MTI simulations
     (\S~\ref{subsec:mti_saturation}).}
   \label{tab:mti_pure}
   \begin{center}
     \begin{tabular}{@{}lccccc}
       \hline
       Name       & $D$ & res & $L/H$ & $\kappa$ & Field Configuration \\
       \hline
         m1       &   2 &  64 & 0.033 &     7.07 &          horizontal \\    
         m2       &   2 &  64 & 0.033 &     7.07 &          vertical   \\    
         m3       &   3 &  64 & 0.033 &     7.07 &          horizontal \\    
         m4       &   3 &  64 & 0.033 &     7.07 &          vertical   \\    
         m5$^{*}$ &   3 & 128 & 0.500 &     0.31 &          horizontal \\   
         m6$^{*}$ &   3 & 128 & 1.400 &     0.35 &          horizontal \\   
       \hline
     \end{tabular}
   \end{center}

   \medskip
   The definitions of $L$, $D$, and $\kappa$ are the same as in
   Table~\ref{tab:hbi_pure}.  All simulations use the local setup
   (eq.~\ref{eq:mti_local_setup}), except for the one with L/H = 1.4, which is
   global (eq.~\ref{eq:global_setup}).  Each of these simulations was
   initialized with a weak magnetic field $B/\sqrt{4\pi} = 10^{-4}$ with the
   orientation indicated in the table.  $^{*}$We also repeated simulations m5
   and m6 with initial field strengths $B/\sqrt{4\pi} = {10^{-4}, 0.0014,
   0.0245}$
 \end{table}

 As we did for the HBI, we study the saturation of the MTI using 2D and 3D
 simulations spanning a range of domain sizes $L/H$.  Table~\ref{tab:mti_pure}
 summarizes the simulations presented in this section.

 Since the linear dispersion relation successfully describes the nonlinear
 evolution and saturation of the HBI, it is a good place to begin our
 discussion of the MTI.  The MTI is described by
 equation~\eqref{eq:simplified_dispersion} when $\partial T / \partial z < 0$.
 The linear evolution of the MTI is the opposite of that of the HBI: the MTI
 operates when the temperature decreases with height, its fastest growing
 modes are the ones with wave vectors $\vec{k}$ parallel to $\hat{b}$, and the
 force that destabilizes the MTI is exactly that which stabilizes the HBI in
 its saturated state.  Equation~\eqref{eq:simplified_dispersion} shows that
 the maximum growth rate of the MTI goes to zero when $\oldhat{b}_z = 1$.  By
 analogy with the HBI, it thus seems reasonable to expect that the MTI also
 saturates quiescently, by making the field lines vertical.

 The first three panels of Figure~\ref{fig:mti_5panel_h} show that this is
 nearly what happens.  As the perturbations grow exponentially, the buoyantly
 rising and sinking blobs rake out the field lines, making them largely
 vertical.  The growth rate of the MTI goes to zero when the field lines
 become vertical; since the velocities are still small at this point in the
 evolution ($\sim 10^{-2} c_s$), one might expect the MTI to operate like the
 HBI and quiescently settle into this stable equilibrium state.  Instead,
 however, the MTI drives sustained turbulence for as long as the temperature
 gradient persists.  The plasma never becomes buoyantly stable, and the
 magnetic field and fluid velocities are nearly isotropic at late times.

 We can understand this evolution using the same approach we employed for the
 HBI.  Although the plasma in our MTI simulations never reaches a state in
 which the MTI growth rate is zero, examining the properties of this state is
 very instructive.  The equilibrium state of the MTI with $\oldhat{b}_z = 1$
 (i.e., a vertical field) has precisely the same dispersion relation as the
 saturated state of the HBI, given by equation~\eqref{eq:hbi_stable_osc}.
 There are again zero frequency (neutrally stable) modes of the dispersion
 relation which correspond to horizontal perturbations to the equilibrium
 state of the MTI; these experience no restoring force, because the restoring
 force is buoyant in nature and unaffected by horizontal
 displacements.\footnote{This conclusion is more subtle than the analogous
 argument for the HBI, because the equilibrium state has a nonzero heat flux.
 If the horizontal displacements aren't incompressive, the field lines could
 pinch together, heating and destabilizing parts of the plasma.  These
 perturbations do not appear in eq~\eqref{eq:hbi_stable_osc} because we have
 taken the Boussinesq limit.  It is in principle possible that such
 compressive perturbations contribute to destabilizing the $\oldhat{b}_z = 1$
 MTI state in our simulations, but all of our analysis is consistent with the
 neutrally stable zero frequency perturbations being the critical ingredient.}
 Critically, however, these zero frequency perturbations now add a horizontal
 component to the magnetic field, pulling the plasma out of the equilibrium
 state and rendering it unstable to the MTI.

 Figure~\ref{fig:mti_5panel_v} vividly illustrates this process.  This figure
 shows a simulation that starts with the linearly {\em stable} equilibrium
 state of the MTI (a vertical magnetic field), seeded with the same highly
 subsonic velocity perturbations as before.  The compressive component of the
 perturbation rapidly damps because the Mach number is small, and buoyancy
 traps the vertical components of the perturbation in small-amplitude
 oscillations, as predicted by equation~\eqref{eq:hbi_stable_osc}.  The
 incompressive, horizontal displacements propagate freely, however, and by $t
 = 7.5 \, t_{\mr{buoy}}$, they have noticeably changed the local orientation
 of the magnetic field.  The plasma is no longer in its stable equilibrium
 state, and by $t = 20 \, t_{\mr{buoy}}$, it is clear that this process has
 excited the MTI.  At late times, the simulations initialized with horizontal
 (linearly unstable; Fig. \ref{fig:mti_5panel_h}) and vertical (linearly
 stable; Fig. \ref{fig:mti_5panel_v}) magnetic fields are qualitatively
 indistinguishable.  Thus, although equation~\eqref{eq:simplified_dispersion}
 shows that a plasma with $\partial T / \partial z < 0$ is linearly stable if
 the magnetic field is vertical, that configuration is nonlinearly unstable.

 The nonlinear instability of the $\oldhat{b}_z = 1$ state of the MTI
 precludes the magnetic saturation channel. The zero frequency horizontal
 motions generate a horizontal magnetic field component from the vertical
 magnetic field, seeding the instability and closing the dynamo loop.  This
 continuously drives the MTI and generates sustained turbulence.  Without a
 linear means to saturate, the MTI grows until nonlinear effects can compete
 with the linear instability, which requires $v \sim c_s$.

\begin{figure}
   \centering
   \includegraphics[width=0.45\textwidth]{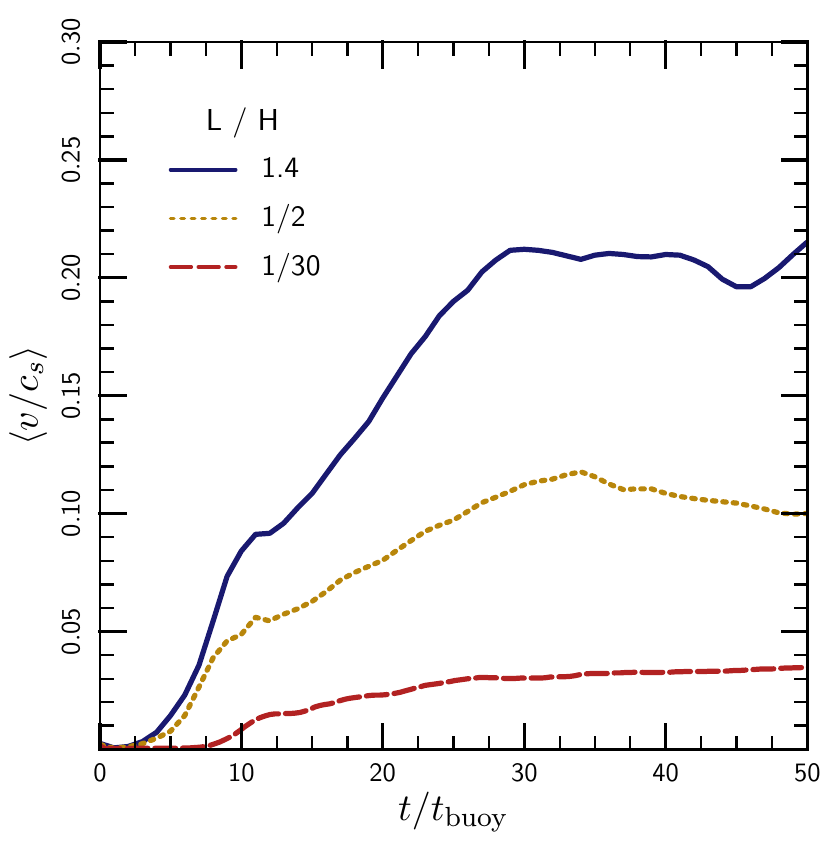}
   \caption{Volume-averaged Mach numbers of the turbulence generated by the
     MTI in 3D simulations, for three different values of the size of
     simulation domain relative to the temperature scale-height $L/H$.  The
     MTI buoyantly accelerates the unstable fluid elements; if the size of the
     simulation domain is smaller than a scale-height, the boundaries suppress
     the growth of the instability. Local simulations with $L/H = 1/30$
     therefore strongly under predict the strength of the turbulence generated
     by the MTI.  In global simulations with $L \gtrsim H$, the MTI leads to
     turbulence with average Mach numbers of $\sim 0.2$; the velocity
     distribution extends up to $\sim 5$ times the mean.}%
   \label{fig:mti_kinetic}
 \end{figure}

 Figure~\ref{fig:mti_kinetic} shows the Mach number in 3D MTI simulations as a
 function of time, for different sizes of the computational domain $L/H$.  The
 MTI buoyantly accelerates rising or sinking fluid elements.  For simulations
 with $L \ll H$, the velocities generated by the MTI are artificially
 suppressed because the small size of the computational domain prematurely
 stops the buoyant acceleration (Fig.~\ref{fig:mti_5panel_h}); the results in
 this Figure~\ref{fig:mti_kinetic} are reasonably consistent with mixing
 length estimate of $v \sim \sqrt{L}$.  By contrast, for simulations with $L
 \gtrsim H$, the Mach numbers approach $\sim 1$ so that the MTI taps into the
 full buoyant force associated with the unstable temperature gradient.  This
 is only true, of course, because our boundary conditions fix the temperature
 at the top and bottom of the computational domain.  If the temperatures were
 free to vary, the MTI could saturate by making the plasma isothermal.  Which
 of these saturation mechanisms is realized in a given astrophysical system
 will depend on the heating and cooling mechanisms that regulate the
 temperature profile of the plasma.

\begin{figure}
   \centering
   \includegraphics[width=0.45\textwidth]{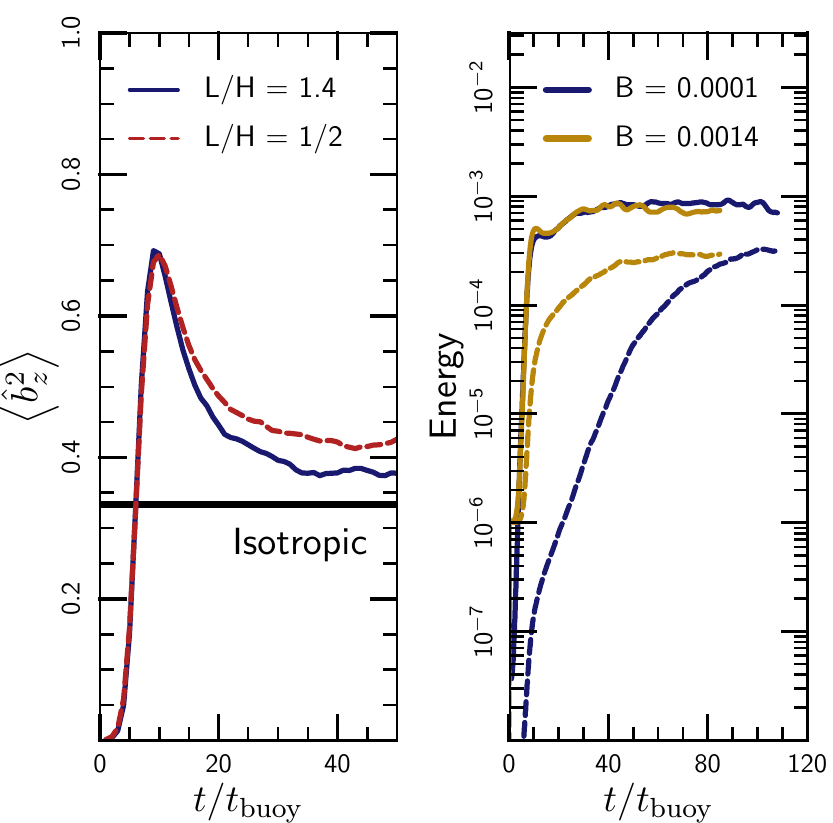}
   \caption{%
     {\bf Left panel}: Evolution of the magnetic field orientation in 3D MTI
     simulations, for two different values of the size of simulation domain
     relative to the temperature scale-height $L/H$. During the linear phase
     of evolution ($t \lesssim 10 \, t_{\mr{buoy}}$), the MTI drives the
     plasma toward a nearly vertical magnetic field, i.e., $\oldhat{b}_z \sim
     1$.  When the instability becomes nonlinear, however, (near $t \sim 10 \,
     t_{\mr{buoy}}$) the evolution changes.  Unlike the HBI, the plasma never
     settles into an equilibrium state; instead the MTI drives vigorous
     turbulence.  This turbulence amplifies and nearly isotropizes the
     magnetic field.
     {\bf Right panel}: Evolution of the magnetic (dashed line) and kinetic
     (solid line) energy in local ($L/H = 1/2$) 3D MTI simulations, with
     different initial field strengths.  These local simulations have a
     positive entropy gradient, but under-predict the magnetic and kinetic
     energy produced by the MTI.  The magnetic energy in the saturated state
     approaches $\sim~10\% \times \rho v^2$.}%
   \label{fig:mti_magnetic}
 \end{figure}

 Figure~\ref{fig:mti_magnetic} shows the evolution of the magnetic field
 orientation (left panel) and energy (right panel) as a function of time in
 our 3D MTI simulations (for two different values of $L/H$).  The sustained,
 vigorous turbulence generated by the MTI in the nonlinear regime ($t \gtrsim
 10 \, t_{\mr{buoy}}$) rapidly amplifies the magnetic field.  The magnetic and
 kinetic energies reach approximate equipartition in our simulations, with
 $B^2/8 \pi \sim 0.1 \, \rho v^2$.  The kinetic and magnetic energies
 generated by the MTI together contribute $\sim$ 5--10\% of the pressure
 support in its saturated state.

 The evolution of the magnetic field geometry, shown in the left panel of
 Figure~\ref{fig:mti_magnetic}, nicely illustrates the transition of the MTI
 from the linear to nonlinear regime.  During the linear phase of the
 instability ($t \lesssim 10 \, t_{\mr{buoy}}$), the plasma accelerates toward
 the equilibrium state with $\oldhat{b}_z = 1$.  After the evolution becomes
 nonlinear ($t \gtrsim 10 \, t_{\mr{buoy}}$), however, the MTI drives
 sustained turbulence, which nearly isotropizes the magnetic field.

\begin{figure}
   \centering
   \includegraphics[width=0.45\textwidth]{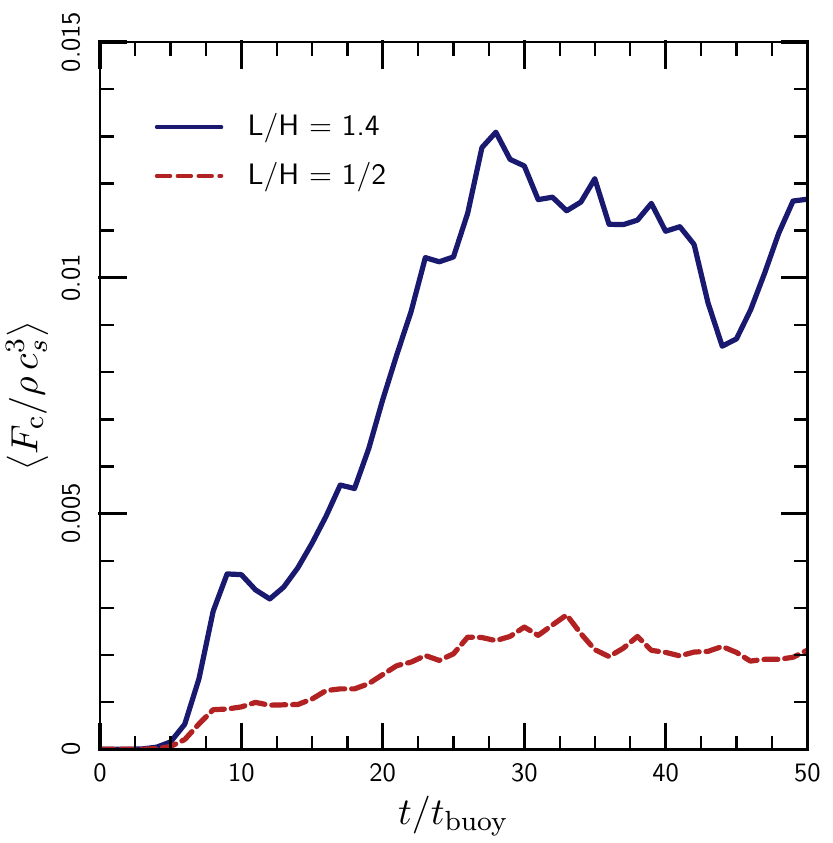}
   \caption{Convective energy fluxes generated by the MTI.  The large
     turbulent velocities associated with the MTI (Fig. \ref{fig:mti_kinetic})
     lead to efficient convective transport of energy, at a reasonable
     fraction of the maximal value, $\rho c_s^3$.  In these MTI simulations,
     the conductive energy flux is $\sim 0.4$ times the field-free value and
     always dominates the convective energy flux for galaxy cluster
     conditions.}%
   \label{fig:mti_fluxes}
 \end{figure}

 While the HBI works to insulate the plasma against vertical energy transport,
 the MTI enhances it.  Figure~\ref{fig:mti_magnetic} shows that the conductive
 flux through the plasma is slightly greater than $\sim 1/3$ of the field free
 value $\kappa_{\mr{e}} \nabla T$ (because $\langle \oldhat{b}_z^2 \rangle
 \sim 0.4$).  Moreover, the MTI leads to large fluid velocities and correlated
 temperature and velocity perturbations---hot pockets of plasma rise, while
 cool pockets sink.  These imply that the MTI drives an efficient outwards
 convective heat flux.  Figure~\ref{fig:mti_fluxes} shows that this flux can
 be $\sim 1.5\%$ of $\rho c_s^3$, consistent with the Mach numbers of $\sim
 0.2$ in Figure~\ref{fig:mti_kinetic}.  This convective flux is probably not
 large enough to influence the thermodynamics of the ICM, but it could be
 important in other environments where the MTI can operate, such as the
 interiors of white dwarfs and neutron stars \citep{Chang2010}.

 %

\section{Interaction with Other Sources of Turbulence}
\label{sec:turbulence}
 Thus far, we have described the development and saturation of buoyancy
 instabilities only in the idealized case of an otherwise quiescent plasma.
 In a more astrophysically realistic scenario, however, other processes and
 sources of turbulence may also act on the plasma and the resulting dynamics
 can be more complicated.  For example, the evolution of the HBI won't simply
 proceed until the growth rate is everywhere zero. Instead, we expect the
 saturated state to involve a statistical balance among the various forces;
 this balance depends on the buoyant properties of the plasma and provides a
 test of our understanding of the nonlinear behavior of the HBI and MTI.
 Furthermore, any change in the saturated state of the plasma due to the
 interaction between the HBI/MTI and other sources of turbulence could change
 the astrophysical implications of these instabilities.

 We choose to explore the interaction between buoyancy instabilities and other
 sources of turbulence using the idealized, isotropic turbulence model
 described in section~\ref{subsec:turbulence}.  While this model glosses over
 the details of what generates the turbulence, we hope that it captures the
 essential physics of the problem, allowing us to study the effect of
 turbulence without unnecessarily restricting our analysis to specific
 applications.  We intend to specialize to specific sources of turbulence in
 future work, but our present analysis should apply in the ICM, accretion
 disks, and anywhere else the assumptions summarized in
 section~\ref{subsec:physics} apply.

 In order to characterize the turbulence, we define a timescale for it to
 influence the plasma.  We define this ``distortion time'' in terms of the
 spatial velocity spectrum: $t_{\mr{dist}}(\ell) = \ell / \delta v (\ell)$,
 where
 \begin{align}
   \delta v (\ell) \equiv \left[ \int
     \left(\delta \vec{v}(\vec{k}) \vphantom{\Sigma^3_1} \right)^2 \;
     \delta \left( | \vec{k} | - 2 \pi/\ell \vphantom{\Sigma^3_1} \right) \;
     \frac{d^3 k}{(2\pi)^3}\right]^{1/2},\label{eq_tdist}
 \end{align}
 and $\delta v(\vec{k})$ is the Fourier transform of the velocity field.  We
 expect the relevant parameter describing the importance of the turbulence to
 be the ratio of the timescales $t_{\mr{buoy}} / t_{\mr{dist}}$.  This
 represents a dimensionless strength of the turbulence; if
 $t_{\mr{buoy}}/t_{\mr{dist}} \gtrsim 1$, the turbulence displaces fluid
 elements faster than buoyancy can restore them, and we expect the velocities
 and magnetic field lines to become isotropic.  In the opposite limit,
 buoyancy still plays an important role in the evolution of the plasma.  The
 scale dependence of the distortion time makes the ratio $t_{\mr{buoy}} /
 t_{\mr{dist}}$ a function of scale.  We define this ratio at the scale where
 the velocity spectrum of the injected turbulence peaks.  This is roughly
 consistent with the driving scale of the turbulence and typically represents
 the scale with the most energy.

 In the following sections, we study the transition from a state dominated by
 buoyancy to one dominated by isotropic turbulence using a number of
 simulations of the HBI and MTI, with turbulence in the range $0.1 \lesssim
 t_{\mr{buoy}} / t_{\mr{dist}} \lesssim 10$.

\subsection{Effect of Turbulence on the HBI}
\label{subsec:hbi_turbulence}
 \begin{table}
   \caption{Parameter study for the HBI simulations with turbulence
     (\S~\ref{subsec:hbi_turbulence}).}
   \label{tab:hbi_turbulence}
   \begin{center}
     \begin{tabular}{@{}ccccccc}
       \hline
       $D$ (2) & res (64) & $L$ (0.1) & $H$ (2.0) & $B_0$  ($10^{-6}$) 
               & $\tilde{k}_0$ \\
       \hline
           --- & --- & --- & --- & ---                &     2 \\
           --- & --- & --- & 1.0 & ---                &     2 \\
           --- & --- & --- & 3.0 & ---                &     2 \\
           --- & --- & --- & --- & ---                &     4 \\
           --- & --- & 1.0 & --- & ---                &     4 \\
           --- & --- & 0.3 & --- & ---                &     4 \\
             3 & --- & --- & --- & ---                &     4 \\
             3 & --- & --- & --- & $10^{-3}$          &     4 \\
             3 & --- & 1.0 & --- & $3 \times 10^{-4}$ &     4 \\
             3 & 128 & 1.0 & --- & $3 \times 10^{-4}$ &     4 \\
           --- & --- & --- & --- & ---                &     6 \\
           --- & --- & --- & --- & ---                &     8 \\
           --- & 256 & --- & --- & ---                &     8 \\
             3 & --- & --- & --- & ---                &     8 \\
       \hline
     \end{tabular}
   \end{center}

   \medskip
   The simulations were initialized with our local setup
   (eq.~\ref{eq:hbi_local_setup}) and an initial magnetic field strength
   $B_0$.  Each simulation was performed on uniform Cartesian grids of side
   $L$, resolution res and dimension $D$.  We varied the size of the
   simulation domain $L$ (scaling the conductivity as described in
   \S~\ref{subsec:global_setup}), the plasma scale height $H$ and the initial
   magnetic field strength $B_0$; the fiducial values for these parameters are
   included in the table header (--- indicates the fiducial value).  For each
   entry in the table, we performed simulations with both initially horizontal
   and vertical magnetic fields.  As described in the
   \S~\ref{subsec:hbi_turbulence}, these simulations include isotropic
   turbulence injected at the scale $k_0 = 2 \pi / L \times \tilde{k}_0$ and
   with a range of turbulent energy injection rates to give $0.1 \lesssim
   t_{\mr{buoy}} / t_{\mr{dist}} \lesssim 10$.
 \end{table}
 %
 \begin{figure}
   \centering
   \includegraphics[width=0.45\textwidth]{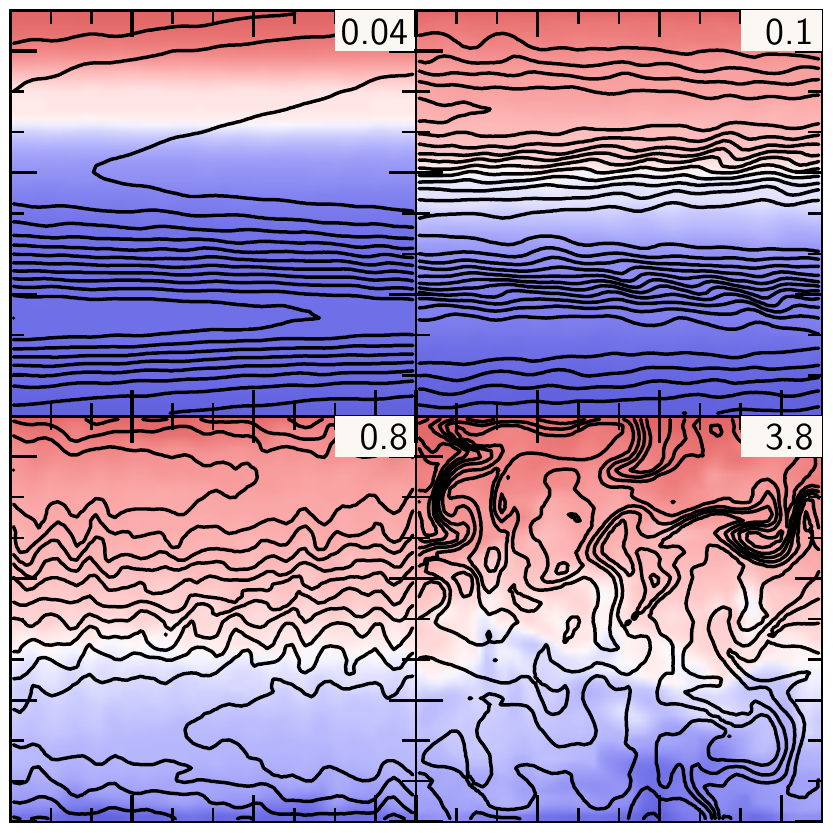}
   \caption{Snapshots of the saturated states of our 2D HBI
     simulations with externally driven turbulence.  Colors show the
     temperature (increasing from blue to red), and black lines show magnetic
     field lines.  Each panel is labeled with the dimensionless strength of
     the turbulence in the simulation, $t_{\mr{buoy}} / t_{\mr{dist}}$,
     defined in \S~\ref{sec:turbulence}.  When $t_{\mr{buoy}} \lesssim
     t_{\mr{dist}}$, as in the top two panels, the HBI dominates the evolution
     of the plasma.  When $t_{\mr{buoy}} \gtrsim t_{\mr{dist}}$, the
     turbulence can isotropize the magnetic field, but it does so in a
     scale-dependent way with the large scales retaining memory of the
     horizontal field imposed by the HBI.}%
   \label{fig:field_lines}
 \end{figure}
 Figure~\ref{fig:field_lines} shows representative snapshots of the
 temperature and magnetic field lines in saturated states of our HBI
 simulations with turbulence; the strength of the injected turbulence
 increases from the top left panel through the bottom right (the labels
 correspond to the values of $t_{\mr{buoy}}/t_{\mr{dist}}$).  When
 $t_{\mr{dist}}$ is long compared to the buoyancy time, as in the first panel
 of Figure~\ref{fig:field_lines}, the turbulence is weak; the HBI therefore
 dominates and the evolution of the plasma is similar to that described in
 section~\ref{subsec:hbi_saturation}.  This saturated state of the HBI feels a
 buoyant restoring force which resists vertical displacements $\xi_z$ with a
 force per unit mass $f_{\mr{buoy}} = \omega_{\mr{buoy}}^2 \xi_z$.  As we
 increase the strength of the applied turbulence in the following panels of
 Figure~\ref{fig:field_lines}, the vertical displacements grow, and the field
 deviates more strongly from the $\oldhat{b}_z = 0$ equilibrium state of the
 HBI.  When $t_{\mr{dist}}$ is short compared to the buoyancy time, as in the
 last panel of Figure~\ref{fig:field_lines}, turbulence can displace the fluid
 elements faster than buoyancy can restore them.  The turbulence then
 dominates the evolution of the plasma, tangling and isotropizing the field
 lines.

 Figure~\ref{fig:field_lines} also shows the length-scale dependence of the
 transition from an HBI to a turbulence dominated state.  It is clear in the
 second and third panels that the HBI has globally rearranged the field lines,
 but that the turbulence is increasingly efficient at smaller scales.  This is
 a consequence of the fact that turbulence typically perturbs the plasma in a
 scale-dependent way, while the buoyant restoring force of the HBI does not.
 If the turbulence follows a Kolmogorov cascade, the force $\sim \omega^2/k
 \propto k^{1/3}$ increases with decreasing scale, so the turbulence will
 always win on sufficiently small length-scales.  It is therefore somewhat
 ambiguous whether turbulence or the HBI dominates a certain configuration, as
 the answer will typically depend on scale.  As mentioned earlier, we skirt
 this issue by defining $t_{\mathrm{dist}}$ at the scale where the velocity
 spectrum of the injected turbulence peaks.  When assessing the astrophysical
 importance of the HBI, it is important to keep this scale in mind.  If the
 scale where the turbulent energy spectrum peaks is smaller than the
 temperature gradient length scale, the HBI may still insulate the plasma
 against conduction, even if the field lines are isotropized on smaller
 scales.

 In order to quantify the transition from an HBI-dominated configuration to
 one dominated by turbulence, we measure the mean orientation of the magnetic
 field via the volume average of $\oldhat{b}_z^2$.  The saturated value for
 this quantity approaches zero when the HBI dominates, and $1/D$ when the
 magnetic field is isotropic, where $D$ is the number of dimensions in the
 simulation.


 \begin{figure}
   \centering
   \includegraphics[width=0.45\textwidth]{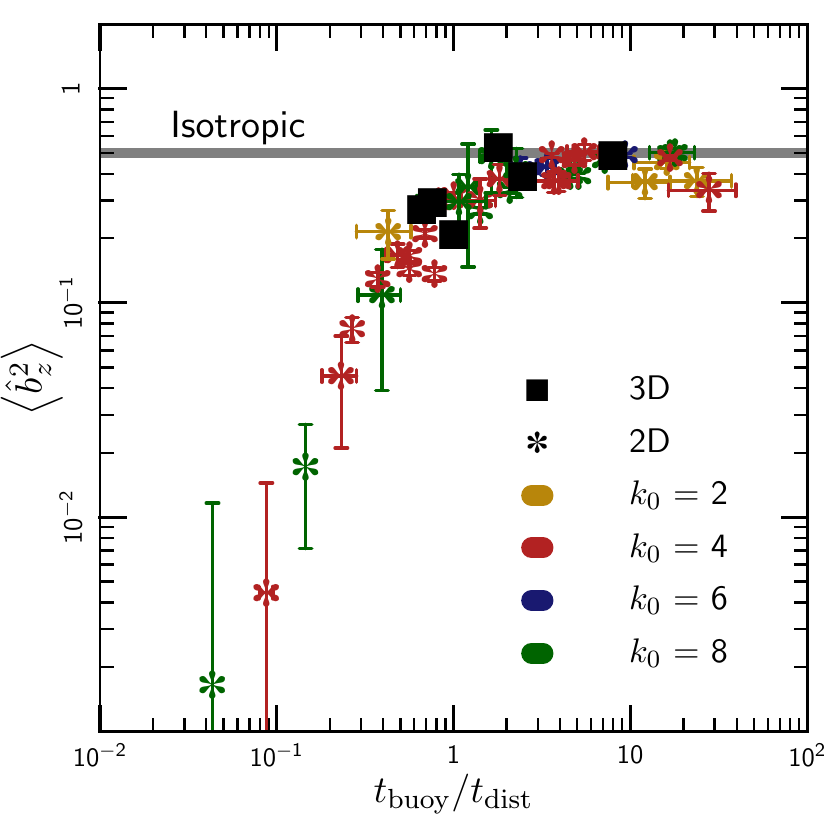}
   \caption{Saturated magnetic field orientation as a function of the
     strength of the externally driven turbulence,
     $t_{\mr{buoy}}/t_{\mr{dist}}$, for local HBI unstable atmospheres with
     $t_{\mr{buoy}} = $~1, $\sqrt{2}$ and $\sqrt{3}$.  The thick gray line is
     an isotropic magnetic field in 2D.  Colored points represent simulations
     with turbulence driven on different scales.  Squares mark 3D
     calculations; the values of $\langle b_z^2 \rangle$ for the 3D
     simulations have been shifted by a factor 3/2 since isotropy implies
     $b_z^2 = 1/2$ in 2D but $b_z^2 = 1/3$ in 3D.  Error bars represent
     $1\sigma$ statistical fluctuations in $b_z$ and $t_{\mr{dist}}$.}%
   \label{fig:saturated_angles}
 \end{figure}
 Figure~\ref{fig:saturated_angles} shows the saturated field angle as a
 function $t_{\mr{buoy}} / t_{\mr{dist}}$.  The points in this figure
 represent simulations with different driving scales $k_0$, different
 dimensionality, and different buoyancy times $t_{\mr{buoy}}$
 (Table~\ref{tab:hbi_turbulence} summarizes our parameter study).  Symmetry of
 the coordinate axes requires that $b_z^2$ = $1/2$ in 2D or $1/3$ in 3D if the
 magnetic field is isotropic.  To include both our 2D and 3D simulations on
 the same plot, we shift our 3D values of $\langle b_z^2 \rangle$ by a factor
 of 3/2.  To within the scatter shown in Figure~\ref{fig:saturated_angles}, we
 find that the saturated value of $\langle b_z^2 \rangle$ depends only on the
 ratio $t_{\mr{buoy}} / t_{\mr{dist}}$: for $t_{\mr{buoy}} \gtrsim
 t_{\mr{dist}}$ the turbulence is strong and the field becomes relatively
 isotropic while for $t_{\mr{buoy}} \lesssim t_{\mr{dist}}$ the isotropic
 turbulence is weak and the HBI drives the magnetic field to become relatively
 horizontal.  The fact that the transition between these two states occurs
 around $t_{\mr{buoy}} \sim t_{\mr{dist}}$ suggests that our definition of
 $t_{\mr{dist}}$, though somewhat arbitrary, is reasonable.

 The bulk of the simulations in Figure~\ref{fig:saturated_angles} are 2D, and
 we do not have any 3D simulations in the very weak turbulence limit.  These
 simulations are computationally expensive, both because of conduction and
 because we have to run for a long time for turbulence and the HBI to reach a
 statistical steady state; using 2D simulations allowed us to explore a larger
 fraction of the interesting parameter space.  While the development of
 turbulence is very different in two and three dimensions, the HBI is
 essentially two-dimensional in nature.  Moreover, the key dynamics governing
 the interaction between the HBI and the turbulence are dominated by the
 energy-containing scale of the turbulence---the precise power-spectrum of the
 fluctuations (which differs in 2D and 3D) is less critical.  Scaling for
 dimension, we find that the saturated states of our 2D and 3D simulations are
 nearly identical.  We thus believe that results in
 Figure~\ref{fig:saturated_angles} in the weak turbulence limit are a good
 description of the magnetic field structure in 3D systems as well.

 \begin{figure}
   \centering
   \includegraphics[width=0.45\textwidth]{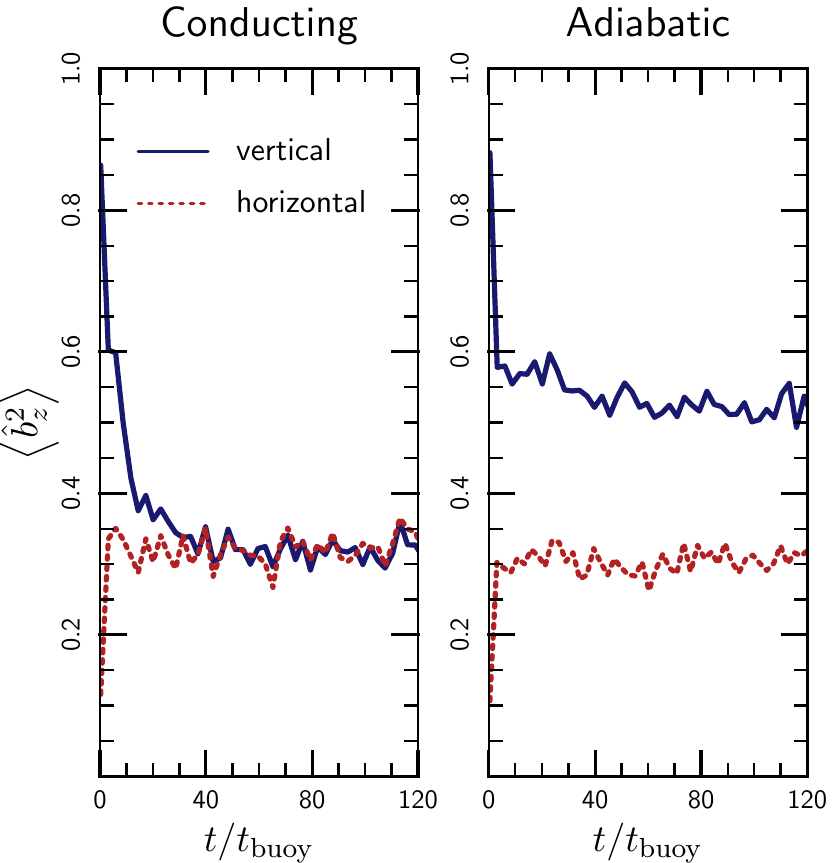}
   \caption{Magnetic field orientation as a function of time in 2D
     simulations with turbulence and the HBI for different initial magnetic
     field orientations; $t_{\mr{buoy}}/t_{\mr{dist}} = 0.8$.  {\em Left
     panel:} Simulations with anisotropic thermal conduction. {\em Right
     panel:} Adiabatic simulations with no conduction (and thus no HBI).
     Different curves represent simulations with initially vertical and
     horizontal magnetic field lines, respectively.  Conducting simulations
     eventually reach the same saturated state independent of the initial
     magnetic field direction.  Adiabatic simulations are very similar to the
     conducting ones if the field is initially horizontal, highlighting the
     fact that in the saturated state of the HBI the plasma is buoyantly
     stable and behaves dynamically like an adiabatic fluid.}%
   \label{fig:initial_conditions}
 \end{figure}

 These results on the interaction between the HBI and other sources of
 turbulence support an analogy between the saturated state of the HBI and
 ordinary, adiabatic stable stratification.  The most important effect of the
 HBI in the saturated states of our simulations is to inhibit mixing in the
 direction of gravity.  Ordinary stable stratification also inhibits vertical
 mixing and, as suggested by \citet{Sharma2009}, our parameter $t_{\mr{buoy}}
 / t_{\mr{dist}}$ is analogous to the Richardson number used in the
 hydrodynamics literature.  To further expand on this analogy,
 Figure~\ref{fig:initial_conditions} shows a comparison of anisotropically
 conducting (left panel) and adiabatic (right panel) simulations with equal
 values of $t_{\mr{buoy}}/t_{\mr{dist}}$.  For the adiabatic simulations, we
 define $t_{\mr{buoy}} = t_{\mr{ad}}$, i.e., using the entropy gradient rather
 than the temperature gradient.  The left panel of
 Figure~\ref{fig:initial_conditions} shows simulations with the same injected
 turbulence, but initialized with vertical or horizontal magnetic field lines.
 In these simulations, the saturated state is independent of the initial
 condition.  That is, the interaction between turbulence and the HBI leads to
 a well defined magnetic field orientation that is independent of the initial
 field direction.

 By contrast, in the adiabatic simulations (right panel of
 Fig. ~\ref{fig:initial_conditions}), the final magnetic field orientation
 depends on the initial field direction.  For an initially vertical field in
 an adiabatic plasma, the turbulence slowly isotropizes the magnetic field
 direction.  However, the adiabatic simulations with initially horizontal
 field lines reach a saturated state that is very similar to that of the HBI
 simulations. In the adiabatic simulations, the magnetic field is essentially
 passive, but it traces the fluid displacements. The stable stratification
 competes with the turbulence and sets a typical scale for vertical
 displacements in the saturated state.  This scale, in turn, determines the
 magnetic field geometry.  The magnetic field plays no dynamical role in this
 process.  The fact that the anisotropically conducting simulations reach the
 same statistical steady state highlights that the saturated state of the HBI
 behaves dynamically very much like an adiabatic, stably stratified, plasma.

 \subsection{Effect of Turbulence on the MTI}
 \label{subsec:mti_turbulence}
\begin{figure}
   \centering
   \includegraphics[width=0.45\textwidth]{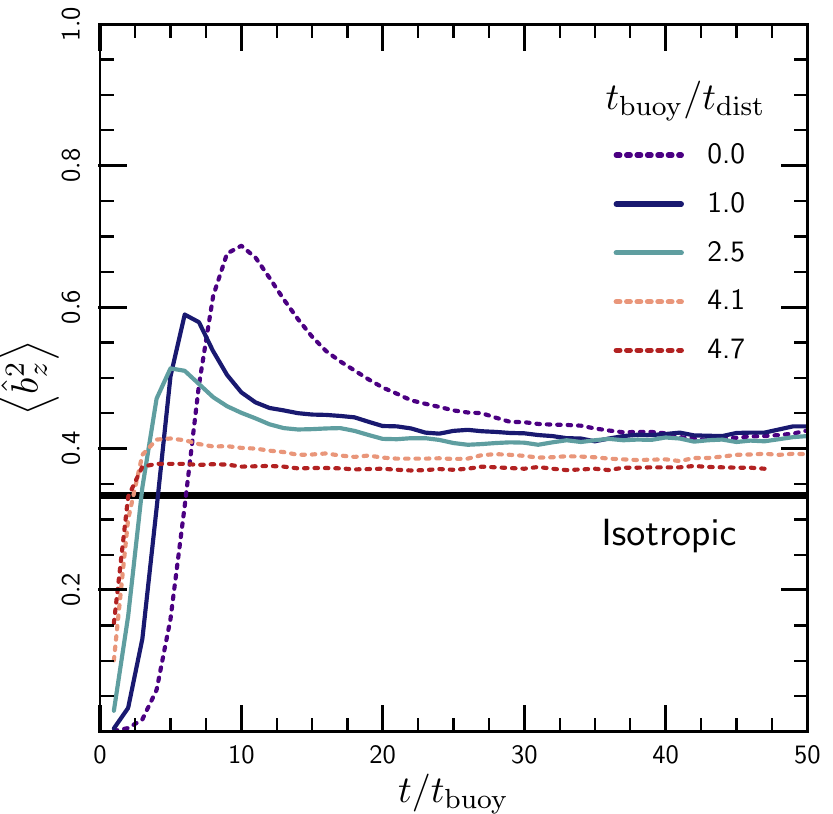}
   \caption{Magnetic field orientation as a function of time in 3D
     simulations with turbulence and the MTI, for different values of the
     strength of the turbulence $t_{\mr{buoy}}/t_{\mr{dist}}$.  In all of the
     simulations, the magnetic field is relatively isotropic in the saturated
     state.  However, the early-time `overshoot' towards a vertical magnetic
     field due to the MTI is suppressed in the presence of strong turbulence
     with $t_{\mr{buoy}} \gtrsim t_{\mr{dist}}$.  These relatively local
     simulations ($L = 0.5 H$) underestimate the kinetic energy generated by
     the MTI (Fig. \ref{fig:mti_kinetic}) and therefore the strength of the
     turbulence required to influence it.}%
   \label{fig:mti_turb_bz}
 \end{figure}

 To complete our analysis, we study how externally imposed isotropic
 turbulence affects the saturation of the MTI.  Figure~\ref{fig:mti_turb_bz}
 shows the volume averaged magnetic field orientation as a function of time in
 MTI simulations with additional turbulence, for different values of the
 strength of the turbulence $t_{\mr{buoy}}/t_{\mr{dist}}$.  These are local
 simulations (with domain sizes $L/H = 0.5$; initialized using
 eq.~\eqref{eq:mti_local_setup}) which have the pedagogical advantage of a
 positive entropy gradient, but which under-predict the kinetic energy
 generated by the MTI.  We drive the turbulence at relatively small scales ($k
 L / 2 \pi = 8$) so that subsonic turbulence can still satisfy $t_{\mr{buoy}}
 / t_{\mr{dist}} > 1$.  Both our driven turbulence and the MTI tend to
 isotropize the magnetic field, so it is not a priori clear whether the field
 orientation is a good indication of the importance of turbulence relative to
 the MTI.  Although Figure~\ref{fig:mti_turb_bz} shows that there is no strong
 dependence of the saturated field orientation on the strength of the
 turbulence, the time dependence of the field orientation clearly shows the
 effects of the turbulence on the MTI.

 In general, the evolution of the MTI proceeds through two stages: there is a
 linear phase, where the plasma accelerates toward its nominal stable state
 (which has a vertical magnetic field), and a nonlinear transition to the
 saturated state, where the strong turbulence generated by the MTI isotropizes
 the velocities and field lines.  In the absence of additional turbulence, the
 linear phase is characterized by field lines that are primarily in the
 direction of gravity (Fig. \ref{fig:mti_5panel_h}).
 Figure~\ref{fig:mti_turb_bz} shows that additional sources of strong (rapidly
 shearing) turbulence suppress this linear phase of the MTI.  Indeed, the
 evolution of the field angle with time in our strongest turbulence
 simulations ($t_{\mr{buoy}}/t_{\mr{dist}} = 4.7$) is quite similar to what we
 find in simulations of an adiabatic plasma in which the MTI is not present.

\begin{figure}
   \centering
   \includegraphics[width=0.45\textwidth]{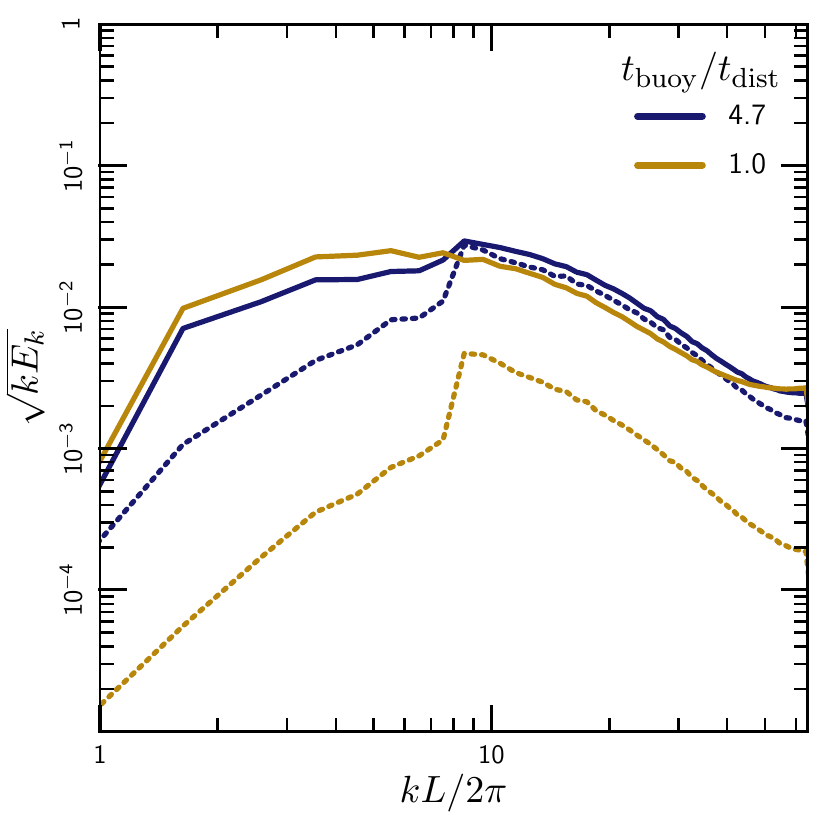}
   \caption{Velocity spectra in simulations with turbulence and
     the MTI (solid lines), for different values of the strength of the
     injected turbulence $t_{\mr{buoy}}/t_{\mr{dist}}$.  The turbulence is
     driven at $k L / 2\pi = 8$. For comparison, dashed lines show the power
     spectra in adiabatic simulations, in which the injected turbulence is the
     only source of power.  Even when $t_{\mr{buoy}} / t_{\mr{dist}} \gg 1$ at
     the driving scale of the turbulence, so that one might expect the MTI to
     be suppressed, the MTI produces significant turbulent energy on larger
     scales $k L / 2\pi \sim 2-10$.  Suppressing the MTI on all scales in a
     plasma would thus require $t_{\mr{buoy}} / t_{\mr{dist}} \gg 1$ on all
     scales smaller than the scale-height.}%
   \label{fig:mti_turb_spectra}
 \end{figure}

 It would, however, be incorrect to conclude from Figure~\ref{fig:mti_turb_bz}
 that the MTI is unimportant if there are other strong sources of turbulence
 in the plasma.  The fundamental reason for this is that the growth of the MTI
 does not depend significantly on scale, while the effects of the other
 sources of turbulence do.  Figure~\ref{fig:mti_turb_spectra} shows velocity
 spectra for the simulations in Figure~\ref{fig:mti_turb_bz}; for comparison
 we also show the velocity spectra in adiabatic simulations, which correspond
 to the power spectra produced solely by the injected turbulence.
 Figure~\ref{fig:mti_turb_bz} demonstrates that even in simulations with very
 strong imposed turbulence ($t_{\mr{buoy}} / t_{\mr{dist}} = 4.7$) there is
 still significant excess power on the largest scales in the computational
 domain ($k L / 2\pi \lesssim 10$). This large-scale power is due to the MTI.
 Moreover, the turbulent energy due to the MTI dominates the total turbulent
 kinetic energy in the plasma.  These results highlight that `strong'
 turbulence is a scale-dependent statement.  Suppressing the MTI requires
 having $t_{\mr{buoy}} / t_{\mr{dist}} \gg 1$ on {\em all scales}, up to the
 temperature/pressure scale-height of the plasma.  Because the MTI itself
 generates nearly sonic velocities, this suppression would require close to
 supersonic turbulence.  In practice, it is therefore unlikely that additional
 sources of turbulence can fully suppress the MTI in most astrophysical
 environments where it is likely to occur (e.g., accretion disks and galaxy
 clusters).

\section{Discussion}
\label{sec:discussion}
 The motion of electrons along, but not across, magnetic field lines in
 dilute, magnetized plasmas produces efficient, anisotropic transport of heat.
 Such plasmas are therefore non-adiabatic, and the standard analysis of
 buoyancy (or convective) instabilities does not necessarily apply.
 Quantitatively, conduction plays an essential role on scales less than $7 \,
 (\lambda H)^{1/2}$, where $\lambda$ is the electron mean free path and $H$ is
 the plasma scale height.  In this ``rapid conduction limit,'' the temperature
 gradient, rather than the entropy gradient, dictates the stability of the
 plasma, and the plasma is unstable for either sign of the temperature
 gradient \citep{Balbus2000, Quataert2008}.  The convective instability in
 this limit is known as the HBI (MTI) when the temperature increases
 (decreases) with height.

 \citet{Parrish2005, Parrish2007} and \citet{Parrish2008hbi} extended the
 original linear analysis of the MTI and HBI into the nonlinear regime using
 numerical simulations.  In this paper, we have reconsidered the nonlinear
 saturation of the HBI and MTI.  Our work adds to previous investigations
 because we have identified a key difference between the two instabilities and
 are able to understand the nonlinear behavior of the MTI more completely.
 This paper therefore represents a significant change in our understanding of
 the possible astrophysical implications of the MTI (but not the HBI).  We
 have also studied the effect of an external source of turbulence on both the
 MTI and HBI.  We conclude that other sources of turbulence in a plasma can
 change the saturation of the HBI, but that it is much harder to disrupt the
 MTI.  Below we summarize our results and discuss their astrophysical
 implications, focusing on the intracluster medium (ICM) of galaxy clusters.

\subsection{HBI}
\label{subsec:hbi_discussion}
 The HBI occurs whenever the temperature increases with height in an
 anisotropically conducting plasma.  Plasmas in the rapid conduction limit are
 in general linearly unstable, unless the magnetic field lines are horizontal
 (eq.~\ref{eq:simplified_dispersion}).  Horizontal field lines represent a
 fixed point in the evolution of the plasma: if the plasma somehow reaches
 such a state (and is not perturbed away by another process), it will remain
 there forever.  A horizontal magnetic field is therefore a natural saturation
 channel for the HBI.

 This ``magnetic'' saturation mechanism can be understood using the linear
 dispersion relation of the plasma (\S~\ref{subsec:hbi_saturation}).  Starting
 from a linearly unstable state, the HBI induces both horizontal and vertical
 motions in the plasma.  As the HBI develops, however, the vertical motions
 become trapped in internal gravity waves.  These waves decay, leaving only
 the horizontal motions at late times; thus, the fluid velocities are very
 anisotropic in the saturated state.  Since the horizontal motions don't incur
 a buoyant response, the horizontal displacements can be very large.  These
 motions stretch out the magnetic field lines, amplifying and reorienting
 them, and drive the plasma towards its stable equilibrium with horizontal
 magnetic field lines.  These horizontal motions therefore drive the nonlinear
 evolution and saturation of the HBI.

 Since the saturation of the HBI is dominated by horizontal displacements, the
 key dynamics all occur at approximately the same height in the
 atmosphere. The saturation of the HBI is thus essentially local in nature.
 We have demonstrated this explicitly by carrying out simulations with
 different domain sizes relative to the plasma scale-height; the results of
 these simulations are very similar (Fig.~\ref{fig:hbi_magnetic}).

 The growth rate of the HBI decreases dramatically as the instability
 progresses.  The HBI therefore saturates quiescently, and the velocities in
 the saturated state are very subsonic (Fig.~\ref{fig:hbi_ke_split}).  The
 saturation is driven by horizontal motions with nearly constant velocities,
 so the nonlinear magnetic field amplification is approximately linear, rather
 than exponential, with time.  These findings are consistent with those of
 \citet{Parrish2008hbi}.  We note that the HBI is very much unlike adiabatic
 convection, which can only saturate by changing the thermal state of the
 plasma and therefore generates vigorous turbulence.  The key difference
 between the HBI and adiabatic convection is that the source of free energy
 for the HBI is a conductive heat flux through the plasma, not merely the
 existence of a temperature gradient.  Since the heat flux can be suppressed
 by rearranging the magnetic field, the HBI has a magnetic saturation channel
 that is not available to adiabatic convection.

 The astrophysical implications of the HBI follow immediately from the nature
 of its saturated state.  By reorienting the magnetic field lines, the HBI
 dramatically reduces the conductive heat flux through the plasma.  The HBI
 should operate in the innermost $\sim$100--200~kpc in the intracluster
 medium of cool-core galaxy clusters, where the observed temperature increases
 outward.  As noted in \citet{Parrish2008hbi}, this is precisely where the
 cooling time of the ICM is shorter than its age; the HBI removes thermal
 conduction as a source of energy for the cores, potentially exacerbating the
 cooling flow problem \citep{Parrish2009}.

 Our results demonstrate that the saturated state of the HBI is buoyantly
 stable.  This may not seem surprising, because it is exactly what one would
 expect if the ICM were adiabatic.  The ICM is not, however, adiabatic, and
 thermal conduction would render it buoyantly neutral to vertical
 displacements if the magnetic field lines were tangled.  The saturated state
 of the HBI is buoyantly stable only because of the nearly horizontal magnetic
 field lines (that are perpendicular to the temperature gradient).  The HBI
 therefore inhibits vertical mixing and allows for the existence of weakly
 damped internal gravity waves in the plasma.

 \citet{Sharma2009} noted that the stable stratification associated with the
 saturated state of the HBI competes with other sources of turbulence in a
 well-defined way.  This competition can be understood using a modified
 Richardson number $t_{\mr{buoy}} / t_{\mr{dist}}$, where $t_{\mr{buoy}}$ is
 the timescale for the HBI to grow and $t_{\mr{dist}}$ is a characteristic
 ``distortion time,'' or ``eddy turnover time,'' of the turbulence.  When
 $t_{\mr{buoy}} \gtrsim t_{\mr{dist}}$, the turbulence can isotropize the
 plasma and remove all traces of the HBI (Fig.~\ref{fig:saturated_angles}).
 When $t_{\mr{buoy}} \lesssim t_{\mr{dist}}$, the saturated state of the
 plasma represents a statistical balance between turbulence and the HBI, with
 the magnetic field becoming more horizontal, and the plasma more
 HBI-dominated, for smaller values of $t_{\mr{buoy}} / t_{\mr{dist}}$.  The
 strength of other sources of turbulence is therefore crucial for
 understanding the astrophysical implications of the HBI.

 Figure~\ref{fig:saturated_angles} provides a very simple mapping between the
 properties of the turbulence and the magnetic field geometry in the plasma.
 Given the strength of the turbulence and the thermal state of the plasma,
 this figure provides a recipe for determining the mean geometry of the
 magnetic field, and therefore the effective conductivity of the plasma.  This
 can be used to interpret observational results or to construct semi-analytic
 models of anisotropic conduction for use in cosmological simulations.

 Turbulence in the ICM is currently poorly constrained, and thus it is
 difficult to determine precisely how important the HBI is for the evolution
 of clusters.  Reasonable estimates suggest that $t_{\mr{buoy}} /
 t_{\mr{dist}} \sim 1$, but more detailed simulations of clusters are required
 to determine this ratio more precisely.  Future observations of the ICM with
 space-based x-ray calorimeters will place observational constraints on the
 level of turbulence.  In addition, Faraday rotation measurements of the ICM
 will measure the orientation of the magnetic fields in clusters and constrain
 the role of the HBI \citep{Bogdanovic2010} (\citet{Pfrommer2010} describe
 another mechanism to measure the magnetic orientation in the ICM).  Even if
 the turbulence is strong ($t_{\mr{buoy}} / t_{\mr{dist}} \gtrsim 1$), the
 driving scale and filling factor of the turbulence may allow the HBI to
 dominate on some scales or at some locations (see Fig.~\ref{fig:field_lines}
 and associated discussion in \S~\ref{subsec:hbi_turbulence}).  As suggested
 by \citet{Parrish2010} and \citet{Ruszkowski2010}, the interaction between
 turbulence and the HBI might be part of a feedback loop for the thermal
 evolution of the ICM.

\subsection{MTI}
\label{subsec:mti_discussion}
 The nonlinear evolution of the MTI is more complex than that of the HBI.
 Just like the HBI, the MTI has linearly stable equilibria, but they are
 transposed: the linearly stable equilibrium states of the MTI have vertical
 field lines.  We have shown, however, that the linearly stable equilibrium
 states turn out to be nonlinearly unstable; i.e., they are unstable to
 perturbations with a finite amplitude.  This nonlinear instability arises
 because neutrally buoyant, horizontal displacements add a horizontal
 component to the magnetic field.  This takes the plasma out of its linearly
 stable state and re-seeds the instability (Fig.~\ref{fig:mti_5panel_v} and
 \S~\ref{subsec:mti_saturation}).  As a result, the linearly stable
 equilibrium states of the MTI do not represent fixed points in the evolution
 of the instability, and the MTI cannot saturate simply by reorienting the
 magnetic field.

 This difference eliminates the quiescent, magnetic saturation channel for the
 MTI and dramatically changes its evolution.  Without a linear means to
 saturate, the instability grows until nonlinear effects dominate, which
 occurs when $v \sim c_s$.  Unlike the HBI, the MTI therefore drives strong
 turbulence and operates as an efficient magnetic dynamo, much more akin to
 adiabatic convection.  The astrophysical implications of the MTI are
 therefore entirely different from those of the HBI.  The kinetic and magnetic
 energy generated by the MTI can contribute a significant (up to ten percent)
 non-thermal pressure support to the plasma in the saturated state.  This is
 consistent with observational constraints on non-thermal pressure support in
 the ICM near the virial radius \citep{George2009}.  This non-thermal pressure
 support may have consequences for mass estimates of clusters, which often
 rely on the assumption of hydrostatic equilibrium with thermal pressure
 support.  Note, in particular, that the MTI is predicted to be present at
 precisely the same radii ($\gtrsim$~the scale radius) to which x-ray and SZ
 mass measurements are most sensitive.

 Because the MTI operates by buoyantly accelerating fluid elements until they
 approach the sound speed, the results of numerical simulations of the MTI are
 sensitive to the size of the computational domain.  The boundaries of the
 domain can artificially suppress this acceleration, and simulations with
 sizes smaller than a scale height under-predict the kinetic energy generated
 by the MTI (this was the case in the original MTI simulations of
 \citealt{Parrish2005, Parrish2007}).  The nonlinear development of the MTI is
 therefore quite sensitive to the global thermal state of the plasma, and an
 understanding of the MTI requires more careful numerical simulations than are
 needed for the HBI.

 The saturated state of the MTI corresponds to a largely isotropic magnetic
 field, with a slight but persistent vertical (or radial) bias; this bias is
 robust even in the presence of other sources of strong turbulence
 (Fig.~\ref{fig:mti_turb_bz}).  We find that the magnetic energy generated by
 the MTI saturates at about 30\% of the kinetic energy
 (Fig.~\ref{fig:mti_magnetic}).  However, it may be difficult to
 observationally distinguish the turbulence generated by the MTI from that
 generated by other processes.

 The large velocities generated by the MTI, along with correlations between
 the temperature and velocity perturbations, imply that the MTI drives a large
 convective heat flux, $\sim 1.5\% \times \rho c_s^3$
 (Fig.~\ref{fig:mti_fluxes}).  While this convective flux is probably too
 small to alter the thermal evolution of the ICM, it could be important in
 higher density astrophysical plasmas, where the electron mean free path is
 smaller and conduction isn't as efficient.

 While the MTI cannot saturate by reorienting the magnetic field, it can
 saturate by making the plasma isothermal.  Simulations with Neumann boundary
 conditions in which the temperature at the boundaries is free to adjust find
 that this is the case; the atmosphere becomes isothermal before the MTI has a
 chance to develop \citep{Parrish2008mti}.  We fixed the temperature at the
 top and bottom boundaries of our simulations; this is partially motivated by
 the fact that many galaxy clusters in the local universe are observed to have
 non-negligible temperature gradients.

 \citet{Sharma2008} carried out numerical simulations of the MTI in spherical
 accretion flows and found nearly radial magnetic fields, with modest
 turbulence.  This quasi-linear saturation of the MTI might seem to contradict
 the results presented in this paper.  Note, however, that in the Bondi inflow
 studied by \citet{Sharma2008}, the plasma undergoes at most $\sim 5-10$~MTI
 growth times before flowing in.  After 10 growth times, our simulations also
 show approximately radial field lines and modest turbulence
 (Fig.~\ref{fig:mti_magnetic}).  Moreover, the simulations of
 \citet{Sharma2008} covered a very large dynamic range in radius and may have
 lacked the resolution to see the full nonlinear development of the MTI.

 The MTI growth time in the outer parts of the ICM is about 1~Gyr.  Although
 our typical MTI simulations take $\sim$10--20 growth times to saturate, this
 does not preclude the importance of the MTI in galaxy clusters.
 Figure~\ref{fig:mti_magnetic} shows that there is a long, linear ramp-up
 phase where the instability grows from the tiny perturbations we apply to the
 nonlinear state.  Astrophysical perturbations are unlikely to be this
 subsonic.  Figure~\ref{fig:mti_turb_bz} shows that the MTI can saturate in
 $\sim$2--5 growth times when subjected to larger perturbations, suggesting
 that the MTI is likely to become nonlinear in the outer parts of clusters.
 Cosmological simulations will be required to fully understand the
 implications of the MTI for galaxy clusters.

\acknowledgments Support for I. J. .P and P. S. was provided by NASA through
Chandra Postdoctoral Fellowship grants PF7-80049 and PF8-90054 awarded by the
Chandra X-Ray Center, which is operated by the Smithsonian Astrophysical
Observatory for NASA under contract NAS8-03060.  M.M. and EQ were supported in
part by NASA Grant NNX10AC95G, NSF-DOE Grant PHY-0812811, and the David and
Lucile Packard Foundation.  Computing time was provided by the National
Science Foundation through the Teragrid resources located at the National
Center for Atmospheric Research and the Pittsburgh Supercomputing Center. \\

\bibliography{buoyancy_saturation}

\end{document}